\documentclass[twocolumn,pra,showpacs,superscriptaddress,amssymb,amsmath,amsmath]{revtex4-1} 
\usepackage{graphicx}
\usepackage{epstopdf}
\usepackage{bm}
\usepackage{hyperref}
\usepackage{comment}
\usepackage{color}
\usepackage{amsmath}
\usepackage{tikz}
\usepackage{enumitem}
\usepackage{bbold}
\usepackage{ulem}

\hypersetup{%
   pdfpagemode=None, 
   pdfstartpage=1,
   pdfmenubar=true,
   pdftoolbar=true,
   colorlinks = true,
   linkcolor=blue,
   citecolor=blue,
   urlcolor=blue,
   bookmarksopen=false
 }

\newcommand*{\la}{\langle}
\newcommand*{\ra}{\rangle}

\begin{document}

\title{Phase Detection with Neural Networks: Interpreting the Black Box}

\author{Anna Dawid}
\affiliation{Faculty of Physics,  University of Warsaw, Pasteura 5, 02-093 Warsaw, Poland}
\affiliation{ICFO-Institut de Ci\`encies Fot\`oniques, The Barcelona Institute of Science and Technology, Av. Carl Friedrich Gauss 3, 08860 Castelldefels (Barcelona), Spain}

\author{Patrick Huembeli}
\affiliation{ICFO-Institut de Ci\`encies Fot\`oniques, The Barcelona Institute of Science and Technology, Av. Carl Friedrich Gauss 3, 08860 Castelldefels (Barcelona), Spain}

\author{Micha\l~Tomza}
\affiliation{Faculty of Physics,  University of Warsaw, Pasteura 5, 02-093 Warsaw, Poland}

\author{Maciej Lewenstein}
\affiliation{ICFO-Institut de Ci\`encies Fot\`oniques, The Barcelona Institute of Science and Technology, Av. Carl Friedrich Gauss 3, 08860 Castelldefels (Barcelona), Spain}
\affiliation{ICREA, Pg.~Llu\'is Campanys 23, 08010 Barcelona, Spain}

\author{Alexandre Dauphin}
\affiliation{ICFO-Institut de Ci\`encies Fot\`oniques, The Barcelona Institute of Science and Technology, Av. Carl Friedrich Gauss 3, 08860 Castelldefels (Barcelona), Spain}

\date{\today}

\begin{abstract}
Neural networks (NNs) usually hinder any insight into the reasoning behind their predictions. 
We demonstrate how influence functions can unravel the black box of NN when trained to predict the phases of the one-dimensional extended spinless Fermi-Hubbard model at half-filling.
Results provide strong evidence that the NN correctly learns an order parameter describing the quantum transition in this model.
We demonstrate that influence functions allow to check that the network, trained to recognize known quantum phases, can predict new unknown ones within the data set.
Moreover, we show they can guide physicists in understanding patterns responsible for the phase transition.
This method requires no \textit{a priori} knowledge on the order parameter, has no dependence on the NN's architecture or the underlying physical model, and is therefore applicable to a broad class of physical models or experimental data.

\end{abstract}


\maketitle
\section{Introduction}
Machine learning (ML) influences everyday life in multiple ways with applications like text and voice recognition software, fingerprint identification, self-driving cars, and many others.
These versatile algorithms, dealing with big and high-dimensional data, also have a noticeable impact on science, which harnessed neural networks (NNs) to solve problems of quantum chemistry, material science, and biology~\cite{Behler07PRL, Ward16, Christiansen18, Wong18}.
Physics is no different in exploring ML methods, encompassed already by astrophysics, high-energy physics, quantum state tomography, and quantum computing \cite{Carleo19RevMod, Naul18, Baldi14, Torlai18NatPhys, Carrasquilla19, Torlai18, Bukov18}.
Especially abundant is the use of ML in phase classification.
It is not surprising if one considers that determining the proper order parameters for unknown transitions is no trivial task, on the verge of being an art.
It includes the search in the exponentially large Hilbert space and the examination of symmetries existing in the system, guided by the intuition and educated guess.
The alternative route was shown, when NNs located the~phase transitions for known models without \textit{a~priori} physical knowledge~\cite{Carrasquilla17NatPhys, Nieuwenburg17NatPhys}.
Numerous works followed, studying classical \cite{Carrasquilla17NatPhys, Schafer19, Tanaka17, Li18, Wang16}, quantum \cite{Nieuwenburg17NatPhys, Liu18, Broecker17, Huembeli19, Chng18, Theveniaut19, Wetzel17a, kottmann2020unsupervised, Vargas18b}, and topological phase transitions~\cite{Deng17, Huembeli18, Zhang18PRL, Tsai19, Greplova19} as well as phases in experimental data \cite{Rem19, Khatami20}.
ML not only finds expected phases but also does it at a much lower computational cost, e.g., using fewer samples or smaller system sizes~\cite{Huembeli19, Theveniaut19}.

On the other hand, there are still open questions of ML struggling with topological models and many-body localization (MBL).
These problems include the need for pre-engineered features~\cite{Kim17, Beach18, Richter18}, disagreement of predicted critical exponents~\cite{Huembeli19}, and high sensitivity to hyperparameters describing the training process~\cite{Theveniaut19}.
Moreover, even in the models described by Landau's theory, so far, these approaches have mostly enabled only the recovery of known phase diagrams or the location of phase transitions in qualitative agreement with more conventional methods based, for instance, on order parameters or theory of finite-size scaling.
Most importantly, however, the resulting models are mostly black boxes, i.e., systems with internal logic not obvious at all to a user~\cite{Guidotti18}.
The missing key element is the model interpretability, i.e., the ability to be explained or presented to a~human in understandable terms~\cite{DoshiVelez17}.
Without this property, we cannot learn anything new from the ML model when applying it to unknown physical systems, nor understand its problems with capturing the topological or MBL signatures.
Physicists have already stressed this need for interpretation, but proposed methods are either restricted to linear and kernel models~\cite{Wang16, Wetzel17a, Ponte17, Zhang19a, Greitemann19, Greitemann20} or the particular NN's architecture~\cite{Wetzel17b, Wetzel20} or require pre-engineering of the data, being as a result specific to both the ML and physical model~\cite{Zhang19b}.

\begin{figure}[t]
\begin{center}
\includegraphics[width=\columnwidth]{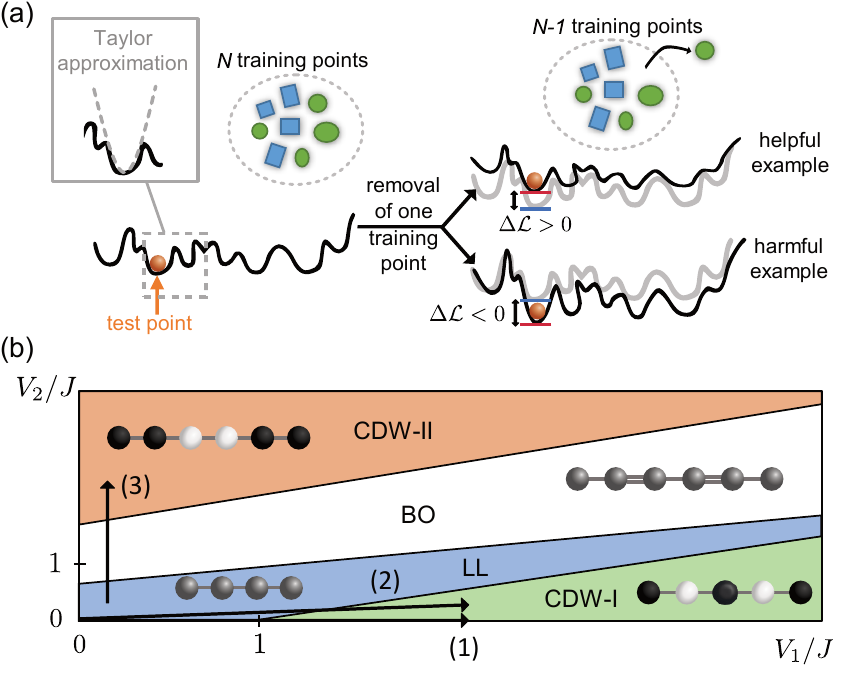}
\end{center}
\vspace{-0.7cm}
\caption{(a) A visual explanation of leave-one-out training and its approximation, the influence function. (b) Schematic phase diagram of the extended one-dimensional half-filled spinless Fermi-Hubbard model with the schemes of the corresponding states: LL - Luttinger liquid, BO - bond order, CDW-I and II - charge density wave type~I and~II. The arrows indicate the transitions studied in this work.}
\label{fig:intro}
\vspace{-0.55cm}
\end{figure}

Hence, in this work, we address the need for interpretability of ML models used in physics on the example of the fundamental one-dimensional (1D) Fermi-Hubbard model.
We follow a paradigm without relying on the \textit{a priori} knowledge on the order parameter or the system itself, with an approach that is straightforwardly applicable to any physical model or experimental data with no dependence on the architecture of the ML model.
We show how the interpretability method, called influence functions, can be used in the quantum phase classification to understand what characteristics are learned by a ML algorithm without, however, providing the order parameter explicitly.
This universal approach unravels if a NN indeed learned a relevant physical concept or cannot be trusted.
We also present how an interpretable NN can give additional information on the transition, not provided to the algorithm explicitly.

\section{Methods}
\subsection{Supervised learning}
We consider supervised learning problems with labeled training data $\mathcal{D} = \{z_i\}_{i=0}^n$, where $z_i = (x_i, y_i)$.
The input data is coming from some input space $x_i \in \mathcal{X}$, and the model predicts the outputs coming from some output space $y_i \in \mathcal{Y}$.
In our setup, the inputs $x_i$ are the state vectors for a given physical system, and $y_i$ are the corresponding phase labels. 
The model is determined by the set of parameters $\theta$.
In the training process, the parameters' space is being searched for the final parameters $\hat{\theta}_{\mathcal{D}} \equiv \hat{\theta}$ of the ML model, which minimize the training loss function $\mathfrak{L}(\mathcal{D}, \theta) = \frac{1}{n} \sum_{z \in \mathcal{D}} \mathcal{L}(z, \theta)$. The training data set size, $n$, tends to be of the order of thousands.
After training, a model can make a prediction for an unseen test point, $z_{\text{test}}$, with the test loss function value, $\mathcal{L}(z_{\text{test}}, \hat{\theta})$, related to the model certainty of this prediction.

\subsection{Interpreting neural networks}
An intuitive way of unraveling the logic learned by the machine is retraining the model after removing a single training point $z_\text{r}$ (starting from the same minimum, if a non-convex problem is analyzed), and checking how it changes the prediction of a specific test point $z_{\text{test}}$.
Such a leave-one-out training (LOO)~\cite{Cook77} studies the change of the parameters $\theta$, now shifted to a new minimum $\hat{\theta}_{\mathcal{D} \setminus \{z_\text{r}\} }$ of the loss function, as depicted in~Fig.~\ref{fig:intro}(a). 
The largest test loss changes, $\Delta \mathcal{L} \equiv \mathcal{L}(z_{\text{test}}, \hat{\theta}) - \mathcal{L}(z_{\text{test}}, \hat{\theta}_{\mathcal{D} \setminus \{z_\text{r}\} })$, indicate the most influential training points for a given test point $z_{\text{test}}$ being the ones whose removal causes the largest change. 
Influential examples can be both helpful ($\Delta \mathcal{L} > 0$) and harmful ($\Delta \mathcal{L} < 0$).
Such an analysis gives the notion of a similarity used by the machine in a given problem, as training points being the closest in the $\Delta \mathcal{L}$ space can be understood as the most similar.
Once the most influential points are indicated, we can decode what characteristics are looked at by comparing 'similar' points in the machine's 'understanding'.
It can be especially useful in phase classification problems where the analysis of $\Delta \mathcal{L}$ enables the recovery of patterns being crucial for distinguishing the phases.
However, this technique is prohibitively expensive, as the model needs retraining for each removed~$z$.

To circumvent this problem, one can make a Taylor expansion of the loss function $\mathcal{L}$ w.r.t. the parameters around the minimum $\hat{\theta}$ and approximate $\Delta \mathcal{L}$ resulting from the LOO training, as presented in~Fig.~\ref{fig:intro}(a). 
This method was proposed for regression problems already forty years ago~\cite{Cook77, Cook80, Cook82} and named influence functions.
This interpretability method is not only computationally feasible but also correctly treats a model as a function of training data.
The influence function reads
\[
\mathcal{I}(z_\text{r}, z_{\text{test}}) =  \frac{1}{n} \nabla_{\theta} \mathcal{L}(z_{\text{test}}, \hat{\theta})^T H^{-1}_{\theta} (\hat{\theta}) \nabla_{\theta} \mathcal{L}(z_\text{r}, \hat{\theta})\,,
\]
and it estimates $\Delta \mathcal{L}$ for a chosen test point $z_{\text{test}}$ after the removal of a chosen training point $z_{\text{r}}$. $\nabla_{\theta} \mathcal{L}(z_\text{test}, \hat{\theta})$ is the gradient of the loss function of the single test point, $\nabla_{\theta} \mathcal{L}(z_\text{r}, \hat{\theta})$ is the gradient of the loss function of the single training point whose removal's impact is being approximated, and $H^{-1}_{\theta} (\hat{\theta})$ is the inverse of Hessian, $H_{i,j} (\hat{\theta}) = \frac{\partial^2}{\partial_{\theta_i} \partial_{\theta_j}} \mathfrak{L}(\mathcal{D}, \theta) |_{\theta = \hat{\theta}}$. 
All derivatives are calculated w.r.t. the model parameters $\theta$, evaluated at $\hat{\theta}$ corresponding to the minimum of the loss, $\mathfrak{L} (\mathcal{D}, \hat{\theta})$.
We can only ensure the existence of the inverse of the Hessian if it is positive-definite. 
\textcolor{black}{It is rarely the case with more sophisticated machine learning models such as neural networks, whose loss landscape is highly non-convex and whose local minima are dominantly flat \cite{Sagun17}.}
However, Koh \textit{et al.} showed that this method can be generalized to such minima and therefore applied to ML \cite{Koh17, Koh19}.
The example code can be found in Ref. \cite{OurRepo}.

\subsection{Physical model} 
We apply influence functions to a small CNN (see Appendix \ref{app:CNN} for the architecture) trained to recognize phases in the extended Hubbard model, namely a 1D system consisting of spinless fermions at half-filling.
The Hubbard models are of fundamental importance to the condensed-matter physics, with the two-dimensional Fermi-Hubbard model believed to describe the high-temperature superconductivity of cuprates~\cite{Dutta15}.
The chosen 1D system has the advantage of being within the power of efficient numerical simulations. As a result, it has a rich and well-studied phase diagram \cite{Hallberg90, Mishra11} and is a promising candidate to be simulated in quantum simulator~\cite{Dutta15}. 
As such, it is suitable to benchmark the influence functions (or any interpretability method) in phase classification problems.
In this model, fermions hop between neighboring sites with amplitudes $ J $ and interact with nearest neighbors with strength $V_1$ and next-nearest neighbors with strength $V_2$
\begin{equation}
\label{eq:ham}
\hat{H} = - J \sum_{\la i,j \ra} c_i^{\dagger} c_j + V_1 \sum_{\la i,j \ra} n_i n_j + V_2 \sum_{\la \la i,j \ra \ra} n_i n_j\,.
\end{equation}
The competition between the system parameters $J$, $V_1$, and $V_2$ leads to four different phases: gapless Luttinger liquid (LL), two gapped charge-density-wave phases with density patterns 1010 (CDW-I) and 11001100 (CDW-II), and bond-order (BO) phase, as seen in~Fig.~\ref{fig:intro}(b). 
The order parameter describing the transition to the CDW-I (-II) phase is the average difference between (next-)nearest-site densities.  Staggered effective hopping amplitudes characterize the BO phase.
We feed the CNN with ground states expressed in the Fock basis, labeled with their appropriate phases, calculated for a 12-site system (see Appendix \ref{app:PD} for the details).
The hopping amplitude, $J$, is set to 1 throughout the paper.

\section{Results}
\subsection{Transition between LL and CDW-I}
\begin{figure}[t]
\begin{center}
\includegraphics[width=\columnwidth]{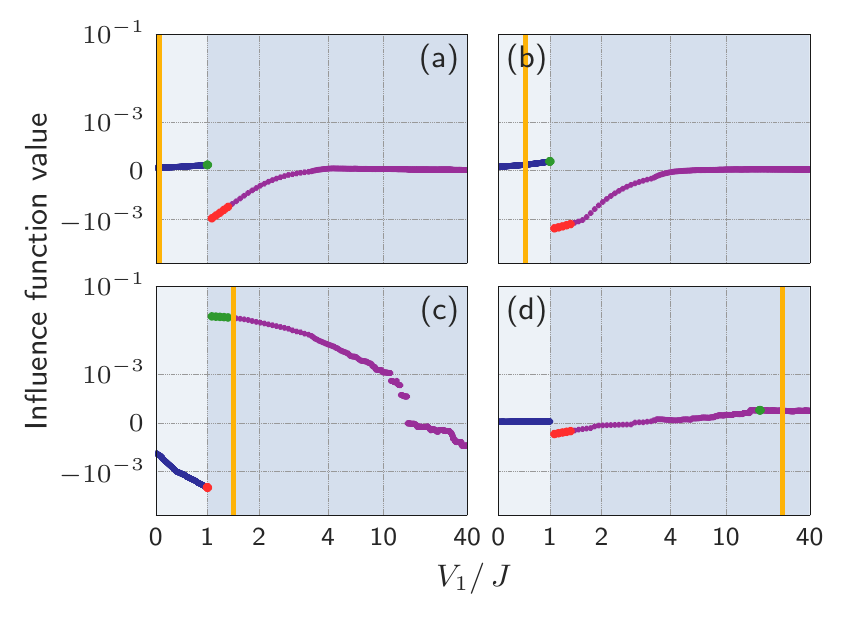}
\end{center}
\vspace{-0.6cm}
\caption{Influence functions of all training examples, i.e., ground states calculated for the transition line between LL and CDW-I for $V_2=0$, marked with dots, for chosen test points marked with an orange line. Blue (purple) dots are influence function values for training examples from the LL (CDW-I) phase. Larger green (red) dots are five the most influential helpful (harmful) training examples. Different background shades indicate two phases. (a)-(b) Blue training points from the LL phase (in blue) are similarly influential to the classification of the test point from the same phase. They all are characterized by a zero order parameter. (c)-(d) The most helpful training examples for the classification of the test points from the CDW-I phase are the ones with the most similar order parameter. Note the use of symmetric log scale.}
\label{fig:J0}
\end{figure}

We train a CNN to classify ground states into two phases: LL and CDW-I based on the transition line marked with the arrow (1) in Fig.~\ref{fig:intro}(b) for $V_2=0$. 
We plot the influence functions of all training examples for a chosen test point (marked with orange line) in Fig.~\ref{fig:J0}.
The order parameter describing the transition here is the average difference between nearest-site densities, which is zero in the LL phase and non-zero (growing to one) in the CDW-I phase.

The panels (a)-(b) present how influential training points are for test points from the LL phase.
The test state (a) is the ground state located deeply in the LL phase, while (b) is closer to the transition.
If the CNN learns an order parameter, all training points, i.e., ground states from the LL phase exhibiting a zero order parameter, should be similarly positively influential, and that is precisely what we observe.
They form an almost flat line in panels (a) and (b).
\textcolor{black}{For both test points (a)-(b) from the LL phase,} the most harmful training points are the ones closest to the transition, but on the CDW-I side. 
These states are the most similar (with the smallest order parameter value), but already labeled differently.

\textcolor{black}{A careful reader can notice that if the CNN learns an order parameter, the training points from the LL phase, all exhibiting a zero order parameter, should be similarly influential and form a flat line in all the panels (a)-(d). However, we see that in reality, their influence changes linearly, what panel (c) shows especially well.
This divergence from expected behavior is mostly due to numerical reasons, and we discuss it in Appendix \ref{app:PD}.}

On the side of the CDW-I phase, the influence pattern is significantly different. 
The curvature of influential points corresponds to the growth of the order parameter, and the most influential helpful points are the ones closest to the test point in the order parameter space, slightly shifted towards the transition point, as they provide more information. 
Panel (c) shows the influence functions of training points for the test states on the CDW-I side, close to the transition. 
The most harmful examples are, as in the previous test points, the ones closest to the transition, but on its other side.
However, panel (d) presents a distinct behavior of the most harmful examples being in the same phase.
All the training points are similarly influential with small values of influence functions resulting in the almost flat line.
It is a signature of the CNN's high certainty regarding the prediction made in panel (d) manifesting with a small test loss function $\mathcal{L}(z_{\text{test}}, \hat{\theta})$.
Also, the analyzed test point is deeply in the CDW-I phase, with all neighboring states being almost identical with the order parameter close to 1.
The most harmful examples are the ones we label as the CDW-I phase, but very different, so the ones closest to the transition.

\textcolor{black}{While analyzing the figures, it is vital to keep in mind that we do not explicitly provide any information on the nearest-neighbor interaction, $V_1\,/J$, present on the $x$-axis (or any physical parameters, in general).
We provide the input states in the random order.
Therefore, the smooth patterns created by the influence functions and resulting ordering of training points, especially on the CDW-I phase's side, is the sole consequence of the internal analysis of the states by the machine.}

\subsection{Transfer learning}
\begin{figure}[t]
\begin{center}
\includegraphics[width=\columnwidth]{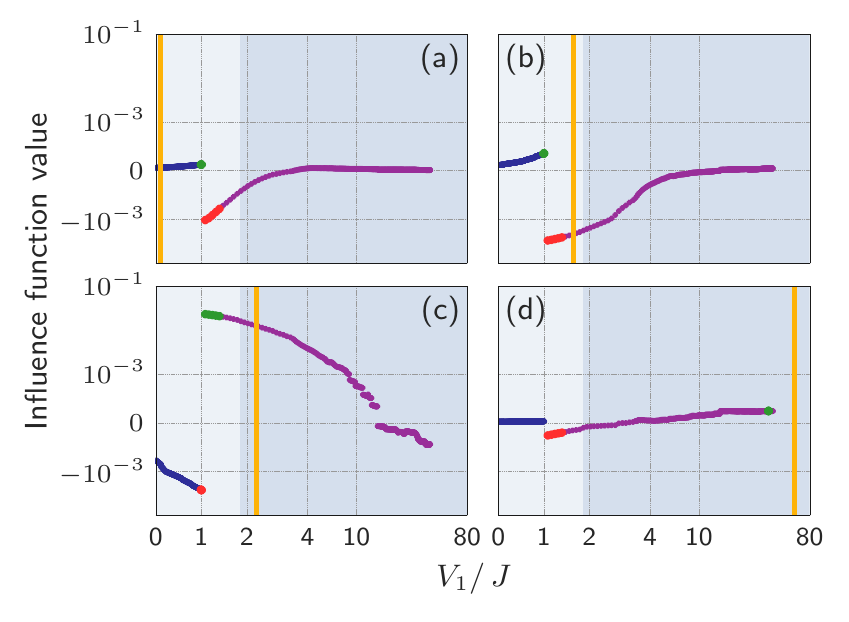}
\end{center}\vspace{-0.6cm}
\caption{Influence functions of all training examples, i.e.,~ground states calculated for the transition line between LL and CDW-I for $V_2=0$, marked with dots, for chosen test states from transition line for $V_2=0.25\, V_1$ marked with an orange line. Blue (purple) dots are influence function values for training examples from the LL (CDW-I) phase for $V_2=0$. Larger green (red) dots are five the most influential helpful (harmful) training examples. Different background shades indicate the phase transition for $V_2=0.25\, V_1$ line, from which test states come from. Panels (a)-(d) show very similar patterns as ones in Fig.~\ref{fig:J0}, but shifted. It indicates that the similarity of test and training points is connected to their order parameters, as the order parameter of test points is shifted towards larger $V_1/\,J$ values, compared to training points, due to coming from the $V_2=0.25\, V_1$ transition line. Note the use of symmetric log scale.}
\label{fig:J025}
\end{figure}

With a similar approach, we validate the transfer learning to another transition line. 
We take the trained CNN from Fig.~\ref{fig:J0}, and in Fig.~\ref{fig:J025} we apply it to test states coming from the transition line for $V_2=0.25 \, V_1$, where the next-nearest-neighbor interaction shifts phase transition to higher values of $V_1/J$.
Therefore the training and test states come from different transition lines, $V_2= 0$ and $0.25 \, V_1$, marked in Fig.~\ref{fig:intro}(b) with the arrows (1) and (2), respectively.
\textcolor{black}{Notice the shift of the panels' backgrounds as compared to Fig.~\ref{fig:J0}. They mark two phases of the test transition line, having a different transition point ($V_1/J = 1.85$) than the training transition line ($V_1/J = 1$).}

Panels (a) and (b) of Fig.~\ref{fig:J025} show the influence function values of training data for test states from the LL phase, while (c) and (d) - from the CDW-I phase.
\textcolor{black}{Panel (a) is identical to the panel (a) of Fig.~\ref{fig:J0}, but already panel (b) shows an interesting divergence from the Fig.~\ref{fig:J0}(b), being a result of a shifted transition point of the test line compared to the training line.
No longer the same value of $V_1/J$, for which test and training states were calculated, yields the same order parameter for both of them.
For example, the test state being in the LL phase, close to the transition point for $V_2=0.25 \, V_1$ should be the most similar to the training points from the LL phase, close to the transition point $V_2=0$, and have the most similar order parameter.
The ML algorithm follows this similarity with regards to the order parameter, what implies a successful transfer learning scheme.
We see similar behavior in panel (c), where the most helpful points are also shifted as compared to Fig.~\ref{fig:J0}(c).
}

\subsection{Inferring the existence of the third phase}
This time we analyze the transition line crossing three phases, LL, BO, and CDW-II, which is indicated by the arrow (3) in Fig.~\ref{fig:intro}(b).
Two order parameters describe this transition.
One is the average difference of the next-nearest neighbor density, which equals zero in the LL and BO phases, and grows to 1 in the CDW-II phase.
The other is the staggering of effective nearest-neighbor hoppings, being 0 in the LL phase, non-zero in the BO phase, and slowly decaying to 0 in the CDW-II phase.
In the studied range of parameters, two phases (BO and CDW-II) co-exist (see Appendix \ref{app:PD} for the details).
It is crucial to note that in this section, we train on the mentioned transition line crossing three phases, but we label ground states only as belonging to one out of two phases.

\begin{figure}[t]
\begin{center}
\includegraphics[width=\columnwidth]{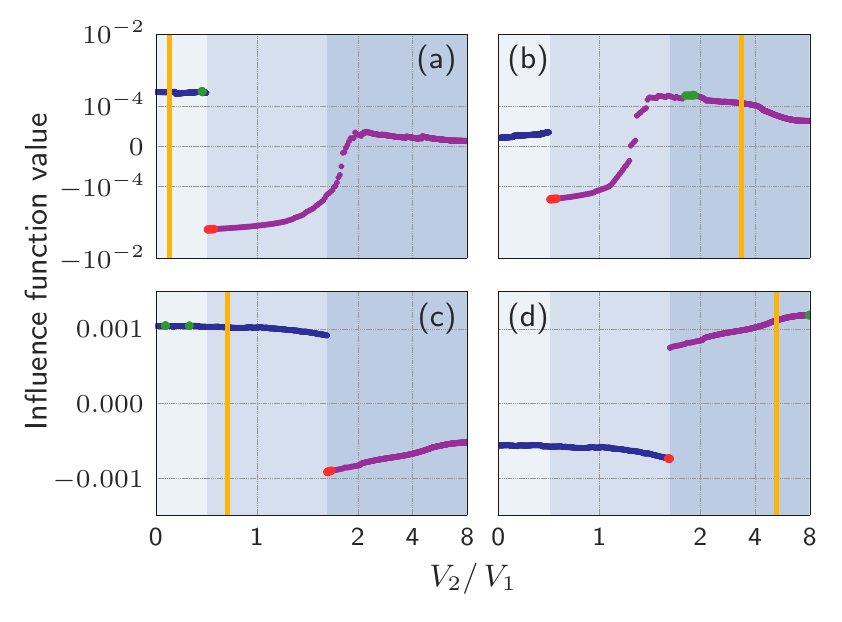}
\end{center}
\vspace{-0.6cm}
\caption{Influence functions of all training examples, i.e.,~ground states calculated for the transition line crossing LL, BO, and CDW-II for $V_1/J = 1$, for chosen test points marked with an orange line. Training examples are marked with dots and labeled differently within two rows, i.e., as (a)-(b) LL - not LL and (c)-(d) CDWII - not CDWII. Blue dots are influence function values for training examples from (a)-(b) the LL phase, (c)-(d) the LL and BO phases, while purple ones (a)-(b) from the BO and CDW-II phases, (c)-(d) from the CDW-II phase. Larger green (red) dots are five the most influential helpful (harmful) training examples. Different background shades indicate phase transitions. (a)-(b) CNN classifies states as belonging to the LL phase or not and detects two similarity regions in the 'not-LL phase'. It effectively indicates the existence of an additional phase. (c)-(d) The model exhibits overfitting. Note the use of a symmetric log scale, except for the linear $y$ axis in panels (c)-(d).}
\label{fig:LL-BO-CDWII}
\end{figure}

In the first set-up, with results presented in the panels (a)-(b) of Fig.~\ref{fig:LL-BO-CDWII}, we label ground states as belonging to the LL (blue dots, label 0) or the BO and CDW-II phases (purple dots, label 1).
Independently on the test point location, notice two similarity regions within purple training points. 
Apparently the NN learns two different patterns (order parameters) to classify the data correctly.
Therefore, it notices the existence of the third phase within the incorrectly labeled data.
Inferring the third phase would be impossible without interpretability methods, which in this sense pave the way towards unknown phases detection.

The second set-up consists of labeling the same data as belonging to the LL and BO phases (blue dots, label 0) or the CDW-II phase (purple dots, label 1). 
The influence functions' values, resulting from this classification, are in the panels (c)-(d) of Fig.~\ref{fig:LL-BO-CDWII}.
The pattern they form is starkly different.
First of all, there is no additional similarity region within training points from the LL and BO.
The behavior is then more similar to the one seen in Fig.~\ref{fig:J0} with the transition between LL and CDW-I.
It is not identical, though, as in the phase LL+BO the most helpful training points are always distributed randomly, but deep in the LL phase, avoiding the BO phase.
The most helpful points on the CDW-II side are deep in the CDW-II phase in contrast to Fig.~\ref{fig:J0}, where they mostly follow the test point.
Consider, that the deeper the CDW-II phase, the smaller the BO order parameter, what makes CDW-II predictions easier.
The observed pattern is the example of NN not learning correctly the order parameter and potentially overfitting.

Finally, we trained a CNN on the same data, but with three labels correctly corresponding to all three phases.
The influence patterns resemble those seen in Fig.~\ref{fig:J0} and panels (c)-(d) of Fig.~\ref{fig:LL-BO-CDWII}, indicating that CNN correctly learns both appropriate order parameters.

\section{Conclusions}
We used the interpretability method called influence functions on the CNN trained in a supervised way to classify ground states of the extended 1D half-filled spinless Fermi-Hubbard model.
We provided strong evidence that the ML algorithm learned a relevant order parameter describing the quantum phase transition.
If no knowledge on the actual order parameter were available, influence functions' values would guide the search for patterns responsible for phase transition and help extract a relevant order parameter, however not providing it explicitly.
We showed that the influence functions, applied to the trained NN, were able to detect an unknown phase.
Two aspects impacted which training points were the most important for a given test point: how similar they were to the test state and how unique within the training data set.
Together they gave a notion of distance or similarity used by the CNN in the phase classification problem and indicated that the patterns relevant for the predictions coincided with the order parameters.

Our approach may be used to address open problems of topological models and MBL with NNs, whose logic can be finally discovered by influence functions.
They may be easily applied to any physical model in general.
\textcolor{black}{Influence functions should be very successful at distinguishing between types of phase transitions.}
\textcolor{black}{In particular, the curvature of the line drawn by influence functions' values should be different for the transitions characterized by continuous and discontinuous change of the order parameter.}
Moreover, this tool proved to be very sensitive to outliers existing in the data set and may serve for anomaly detection.
Finally, along with unsupervised learning techniques, it can serve as the first search for unknown phases and order parameters in experimental data.

\begin{acknowledgments}
\textit{Acknowledgments - }An.D. acknowledges the financial support from the National Science Centre, Poland, within the Preludium grant No. 2019/33/N/ST2/03123. This project has also received funding from the European Unions Horizon 2020 research and innovation programme under the Marie Skłodowska-Curie grant agreement No. 665884 (P.H.).  M.T. acknowledges the financial support from the Foundation for Polish Science within the Homing and First Team programmes co-financed by the EU Regional Development Fund. We (M.L. group) also acknowledge the Spanish Ministry of Economy and Competitiveness (Plan National FISICATEAMO and FIDEUA PID2019-106901GB-I00/10.13039 / 501100011033, “Severo Ochoa” program for Centres of Excellence in R\&D (CEX2019-000910-S), FPI, FIS2020-TRANQI), European Social Fund, Fundació Privada Cellex, Fundació Mir-Puig, Generalitat de Catalunya (AGAUR Grant No. 2017 SGR 1341, CERCA program, QuantumCAT U16-011424, co-funded by ERDF Operational Program of Catalonia 2014-2020), ERC AdG NOQIA, MINECO-EU QUANTERA MAQS (funded by State Research Agency (AEI) PCI2019-111828-2 / 10.13039/501100011033), and the National Science Centre, Poland-Symfonia Grant No. 2016/20/W/ST4/00314. Al.D. acknowledges the Juan de la Cierva program (IJCI-2017-33180) and the financial support from a fellowship granted by la Caixa Foundation (ID 100010434, fellowship code LCF/BQ/PR20/11770012).
\end{acknowledgments}

------------------------------------------------------------------------------
\normalem

\begin{thebibliography}{56}%
\makeatletter
\providecommand \@ifxundefined [1]{%
 \@ifx{#1\undefined}
}%
\providecommand \@ifnum [1]{%
 \ifnum #1\expandafter \@firstoftwo
 \else \expandafter \@secondoftwo
 \fi
}%
\providecommand \@ifx [1]{%
 \ifx #1\expandafter \@firstoftwo
 \else \expandafter \@secondoftwo
 \fi
}%
\providecommand \natexlab [1]{#1}%
\providecommand \enquote  [1]{``#1''}%
\providecommand \bibnamefont  [1]{#1}%
\providecommand \bibfnamefont [1]{#1}%
\providecommand \citenamefont [1]{#1}%
\providecommand \href@noop [0]{\@secondoftwo}%
\providecommand \href [0]{\begingroup \@sanitize@url \@href}%
\providecommand \@href[1]{\@@startlink{#1}\@@href}%
\providecommand \@@href[1]{\endgroup#1\@@endlink}%
\providecommand \@sanitize@url [0]{\catcode `\\12\catcode `\$12\catcode
  `\&12\catcode `\#12\catcode `\^12\catcode `\_12\catcode `\%12\relax}%
\providecommand \@@startlink[1]{}%
\providecommand \@@endlink[0]{}%
\providecommand \url  [0]{\begingroup\@sanitize@url \@url }%
\providecommand \@url [1]{\endgroup\@href {#1}{\urlprefix }}%
\providecommand \urlprefix  [0]{URL }%
\providecommand \Eprint [0]{\href }%
\providecommand \doibase [0]{http://dx.doi.org/}%
\providecommand \selectlanguage [0]{\@gobble}%
\providecommand \bibinfo  [0]{\@secondoftwo}%
\providecommand \bibfield  [0]{\@secondoftwo}%
\providecommand \translation [1]{[#1]}%
\providecommand \BibitemOpen [0]{}%
\providecommand \bibitemStop [0]{}%
\providecommand \bibitemNoStop [0]{.\EOS\space}%
\providecommand \EOS [0]{\spacefactor3000\relax}%
\providecommand \BibitemShut  [1]{\csname bibitem#1\endcsname}%
\let\auto@bib@innerbib\@empty
\bibitem [{\citenamefont {Behler}\ and\ \citenamefont
  {Parrinello}(2007)}]{Behler07PRL}%
  \BibitemOpen
  \bibfield  {author} {\bibinfo {author} {\bibfnamefont {J.}~\bibnamefont
  {Behler}}\ and\ \bibinfo {author} {\bibfnamefont {M.}~\bibnamefont
  {Parrinello}},\ }\bibfield  {title} {\bibinfo {title} {\emph {Generalized
  Neural-Network Representation of High-Dimensional Potential-Energy
  Surfaces}},\ }\href {\doibase 10.1103/PhysRevLett.98.146401} {\bibfield
  {journal} {\bibinfo  {journal} {Phys. Rev. Lett.}\ }\textbf {\bibinfo
  {volume} {98}},\ \bibinfo {pages} {146401} (\bibinfo {year}
  {2007})}\BibitemShut {NoStop}%
\bibitem [{\citenamefont {Ward}\ and\ \citenamefont
  {Wolverton}(2016)}]{Ward16}%
  \BibitemOpen
  \bibfield  {author} {\bibinfo {author} {\bibfnamefont {L.}~\bibnamefont
  {Ward}}\ and\ \bibinfo {author} {\bibfnamefont {C.}~\bibnamefont
  {Wolverton}},\ }\bibfield  {title} {\bibinfo {title} {\emph {Atomistic
  calculations and materials informatics: A review}},\ }\href {\doibase
  10.1016/j.cossms.2016.07.002} {\bibfield  {journal} {\bibinfo  {journal}
  {Curr. Opin. Solid State Mater. Sci.}\ }\textbf {\bibinfo {volume} {21}},\
  \bibinfo {pages} {167} (\bibinfo {year} {2016})}\BibitemShut {NoStop}%
\bibitem [{\citenamefont {{E. M. Christiansen \textit{et
  al.}}}(2018)}]{Christiansen18}%
  \BibitemOpen
  \bibfield  {author} {\bibinfo {author} {\bibnamefont {{E. M. Christiansen
  \textit{et al.}}}},\ }\bibfield  {title} {\bibinfo {title} {\emph {\textit{In
  silico} Labeling: Predicting Fluorescent Labels in Unlabeled Images}},\
  }\href {\doibase 10.1016/j.cell.2018.03.040} {\bibfield  {journal} {\bibinfo
  {journal} {Cell}\ }\textbf {\bibinfo {volume} {173}},\ \bibinfo {pages} {792}
  (\bibinfo {year} {2018})}\BibitemShut {NoStop}%
\bibitem [{\citenamefont {Wong}\ and\ \citenamefont {Yip}(2018)}]{Wong18}%
  \BibitemOpen
  \bibfield  {author} {\bibinfo {author} {\bibfnamefont {D.}~\bibnamefont
  {Wong}}\ and\ \bibinfo {author} {\bibfnamefont {S.}~\bibnamefont {Yip}},\
  }\bibfield  {title} {\bibinfo {title} {\emph {Machine learning classifies
  cancer}},\ }\href {\doibase 10.1038/d41586-018-02881-7} {\bibfield  {journal}
  {\bibinfo  {journal} {Nature}\ }\textbf {\bibinfo {volume} {555}},\ \bibinfo
  {pages} {446} (\bibinfo {year} {2018})}\BibitemShut {NoStop}%
\bibitem [{\citenamefont {Carleo}\ \emph {et~al.}(2019)\citenamefont {Carleo},
  \citenamefont {Cirac}, \citenamefont {Cranmer}, \citenamefont {Daudet},
  \citenamefont {Schuld}, \citenamefont {Tishby}, \citenamefont
  {Vogt-Maranto},\ and\ \citenamefont {Zdeborov\'{a}}}]{Carleo19RevMod}%
  \BibitemOpen
  \bibfield  {author} {\bibinfo {author} {\bibfnamefont {G.}~\bibnamefont
  {Carleo}}, \bibinfo {author} {\bibfnamefont {I.}~\bibnamefont {Cirac}},
  \bibinfo {author} {\bibfnamefont {K.}~\bibnamefont {Cranmer}}, \bibinfo
  {author} {\bibfnamefont {L.}~\bibnamefont {Daudet}}, \bibinfo {author}
  {\bibfnamefont {M.}~\bibnamefont {Schuld}}, \bibinfo {author} {\bibfnamefont
  {N.}~\bibnamefont {Tishby}}, \bibinfo {author} {\bibfnamefont
  {L.}~\bibnamefont {Vogt-Maranto}}, \ and\ \bibinfo {author} {\bibfnamefont
  {L.}~\bibnamefont {Zdeborov\'{a}}},\ }\bibfield  {title} {\bibinfo {title}
  {\emph {Machine learning and the physical sciences}},\ }\href {\doibase
  10.1103/RevModPhys.91.045002} {\bibfield  {journal} {\bibinfo  {journal}
  {Rev. Mod. Phys.}\ }\textbf {\bibinfo {volume} {91}},\ \bibinfo {pages}
  {045002} (\bibinfo {year} {2019})}\BibitemShut {NoStop}%
\bibitem [{\citenamefont {Naul}\ \emph {et~al.}(2018)\citenamefont {Naul},
  \citenamefont {Bloom}, \citenamefont {P\'{e}rez},\ and\ \citenamefont
  {van~der Walt}}]{Naul18}%
  \BibitemOpen
  \bibfield  {author} {\bibinfo {author} {\bibfnamefont {B.}~\bibnamefont
  {Naul}}, \bibinfo {author} {\bibfnamefont {J.~S.}\ \bibnamefont {Bloom}},
  \bibinfo {author} {\bibfnamefont {F.}~\bibnamefont {P\'{e}rez}}, \ and\
  \bibinfo {author} {\bibfnamefont {S.}~\bibnamefont {van~der Walt}},\
  }\bibfield  {title} {\bibinfo {title} {\emph {A recurrent neural network for
  classification of unevenly sampled variable stars}},\ }\href {\doibase
  10.1038/s41550-017-0321-z} {\bibfield  {journal} {\bibinfo  {journal} {Nat.
  Astron.}\ }\textbf {\bibinfo {volume} {2}},\ \bibinfo {pages} {151} (\bibinfo
  {year} {2018})}\BibitemShut {NoStop}%
\bibitem [{\citenamefont {Baldi}\ \emph {et~al.}(2014)\citenamefont {Baldi},
  \citenamefont {Sadowski},\ and\ \citenamefont {Whiteson}}]{Baldi14}%
  \BibitemOpen
  \bibfield  {author} {\bibinfo {author} {\bibfnamefont {P.}~\bibnamefont
  {Baldi}}, \bibinfo {author} {\bibfnamefont {P.}~\bibnamefont {Sadowski}}, \
  and\ \bibinfo {author} {\bibfnamefont {D.}~\bibnamefont {Whiteson}},\
  }\bibfield  {title} {\bibinfo {title} {\emph {Searching for exotic particles
  in high-energy physics with deep learning}},\ }\href {\doibase
  10.1038/ncomms5308} {\bibfield  {journal} {\bibinfo  {journal} {Nat.
  Commun.}\ }\textbf {\bibinfo {volume} {5}},\ \bibinfo {pages} {4308}
  (\bibinfo {year} {2014})}\BibitemShut {NoStop}%
\bibitem [{\citenamefont {Torlai}\ \emph {et~al.}(2018)\citenamefont {Torlai},
  \citenamefont {Mazzola}, \citenamefont {Carrasquilla}, \citenamefont
  {Troyer}, \citenamefont {Melko},\ and\ \citenamefont
  {Carleo}}]{Torlai18NatPhys}%
  \BibitemOpen
  \bibfield  {author} {\bibinfo {author} {\bibfnamefont {G.}~\bibnamefont
  {Torlai}}, \bibinfo {author} {\bibfnamefont {G.}~\bibnamefont {Mazzola}},
  \bibinfo {author} {\bibfnamefont {J.}~\bibnamefont {Carrasquilla}}, \bibinfo
  {author} {\bibfnamefont {M.}~\bibnamefont {Troyer}}, \bibinfo {author}
  {\bibfnamefont {R.}~\bibnamefont {Melko}}, \ and\ \bibinfo {author}
  {\bibfnamefont {G.}~\bibnamefont {Carleo}},\ }\bibfield  {title} {\bibinfo
  {title} {\emph {Neural-network quantum state tomography}},\ }\href {\doibase
  10.1038/s41567-018-0048-5} {\bibfield  {journal} {\bibinfo  {journal} {Nat.
  Phys.}\ }\textbf {\bibinfo {volume} {14}},\ \bibinfo {pages} {447} (\bibinfo
  {year} {2018})}\BibitemShut {NoStop}%
\bibitem [{\citenamefont {Carrasquilla}\ \emph {et~al.}(2019)\citenamefont
  {Carrasquilla}, \citenamefont {Torlai},\ and\ \citenamefont
  {Melko}}]{Carrasquilla19}%
  \BibitemOpen
  \bibfield  {author} {\bibinfo {author} {\bibfnamefont {J.}~\bibnamefont
  {Carrasquilla}}, \bibinfo {author} {\bibfnamefont {G.}~\bibnamefont
  {Torlai}}, \ and\ \bibinfo {author} {\bibfnamefont {R.~G.}\ \bibnamefont
  {Melko}},\ }\bibfield  {title} {\bibinfo {title} {\emph {Latent Space
  Purification via Neural Density Operators}},\ }\href {\doibase
  10.1038/s42256-019-0028-1} {\bibfield  {journal} {\bibinfo  {journal} {Nat.
  Mach. Intell.}\ }\textbf {\bibinfo {volume} {1}},\ \bibinfo {pages} {155}
  (\bibinfo {year} {2019})}\BibitemShut {NoStop}%
\bibitem [{\citenamefont {Torlai}\ and\ \citenamefont
  {Melko}(2018)}]{Torlai18}%
  \BibitemOpen
  \bibfield  {author} {\bibinfo {author} {\bibfnamefont {G.}~\bibnamefont
  {Torlai}}\ and\ \bibinfo {author} {\bibfnamefont {R.~G.}\ \bibnamefont
  {Melko}},\ }\bibfield  {title} {\bibinfo {title} {\emph {Latent Space
  Purification via Neural Density Operators}},\ }\href {\doibase
  10.1103/PhysRevLett.120.240503} {\bibfield  {journal} {\bibinfo  {journal}
  {Phys. Rev. Lett.}\ }\textbf {\bibinfo {volume} {120}},\ \bibinfo {pages}
  {240503} (\bibinfo {year} {2018})}\BibitemShut {NoStop}%
\bibitem [{\citenamefont {Bukov}\ \emph {et~al.}(2018)\citenamefont {Bukov},
  \citenamefont {Day}, \citenamefont {Sels}, \citenamefont {Weinberg},
  \citenamefont {Polkovnikov},\ and\ \citenamefont {Mehta}}]{Bukov18}%
  \BibitemOpen
  \bibfield  {author} {\bibinfo {author} {\bibfnamefont {M.}~\bibnamefont
  {Bukov}}, \bibinfo {author} {\bibfnamefont {A.~G.~R.}\ \bibnamefont {Day}},
  \bibinfo {author} {\bibfnamefont {D.}~\bibnamefont {Sels}}, \bibinfo {author}
  {\bibfnamefont {P.}~\bibnamefont {Weinberg}}, \bibinfo {author}
  {\bibfnamefont {A.}~\bibnamefont {Polkovnikov}}, \ and\ \bibinfo {author}
  {\bibfnamefont {P.}~\bibnamefont {Mehta}},\ }\bibfield  {title} {\bibinfo
  {title} {\emph {Reinforcement Learning in Different Phases of Quantum
  Control}},\ }\href {\doibase 10.1103/PhysRevX.8.031086} {\bibfield  {journal}
  {\bibinfo  {journal} {Phys. Rev. X}\ }\textbf {\bibinfo {volume} {8}},\
  \bibinfo {pages} {031086} (\bibinfo {year} {2018})}\BibitemShut {NoStop}%
\bibitem [{\citenamefont {Carrasquilla}\ and\ \citenamefont
  {Melko}(2017)}]{Carrasquilla17NatPhys}%
  \BibitemOpen
  \bibfield  {author} {\bibinfo {author} {\bibfnamefont {J.}~\bibnamefont
  {Carrasquilla}}\ and\ \bibinfo {author} {\bibfnamefont {R.~G.}\ \bibnamefont
  {Melko}},\ }\bibfield  {title} {\bibinfo {title} {\emph {Machine learning
  phases of matter}},\ }\href {\doibase 10.1038/nphys4035} {\bibfield
  {journal} {\bibinfo  {journal} {Nat. Phys.}\ }\textbf {\bibinfo {volume}
  {13}},\ \bibinfo {pages} {431} (\bibinfo {year} {2017})}\BibitemShut
  {NoStop}%
\bibitem [{\citenamefont {van Nieuwenburg}\ \emph {et~al.}(2017)\citenamefont
  {van Nieuwenburg}, \citenamefont {Liu},\ and\ \citenamefont
  {Huber}}]{Nieuwenburg17NatPhys}%
  \BibitemOpen
  \bibfield  {author} {\bibinfo {author} {\bibfnamefont {E.~P.~L.}\
  \bibnamefont {van Nieuwenburg}}, \bibinfo {author} {\bibfnamefont {Y.-H.}\
  \bibnamefont {Liu}}, \ and\ \bibinfo {author} {\bibfnamefont {S.~D.}\
  \bibnamefont {Huber}},\ }\bibfield  {title} {\bibinfo {title} {\emph
  {Learning phase transitions by confusion}},\ }\href {\doibase
  10.1038/nphys4037} {\bibfield  {journal} {\bibinfo  {journal} {Nat. Phys.}\
  }\textbf {\bibinfo {volume} {13}},\ \bibinfo {pages} {435} (\bibinfo {year}
  {2017})}\BibitemShut {NoStop}%
\bibitem [{\citenamefont {Sch\"afer}\ and\ \citenamefont
  {L\"orch}(2019)}]{Schafer19}%
  \BibitemOpen
  \bibfield  {author} {\bibinfo {author} {\bibfnamefont {F.}~\bibnamefont
  {Sch\"afer}}\ and\ \bibinfo {author} {\bibfnamefont {N.}~\bibnamefont
  {L\"orch}},\ }\bibfield  {title} {\bibinfo {title} {\emph {Vector field
  divergence of predictive model output as indication of phase transitions}},\
  }\href {\doibase 10.1103/PhysRevE.99.062107} {\bibfield  {journal} {\bibinfo
  {journal} {Phys. Rev. E}\ }\textbf {\bibinfo {volume} {99}},\ \bibinfo
  {pages} {062107} (\bibinfo {year} {2019})}\BibitemShut {NoStop}%
\bibitem [{\citenamefont {Tanaka}\ and\ \citenamefont
  {Tomiya}(2017)}]{Tanaka17}%
  \BibitemOpen
  \bibfield  {author} {\bibinfo {author} {\bibfnamefont {A.}~\bibnamefont
  {Tanaka}}\ and\ \bibinfo {author} {\bibfnamefont {A.}~\bibnamefont
  {Tomiya}},\ }\bibfield  {title} {\bibinfo {title} {\emph {Detection of Phase
  Transition via Convolutional Neural Networks}},\ }\href {\doibase
  10.7566/JPSJ.86.063001} {\bibfield  {journal} {\bibinfo  {journal} {J. Phys.
  Soc. Jpn.}\ }\textbf {\bibinfo {volume} {86}},\ \bibinfo {pages} {063001}
  (\bibinfo {year} {2017})}\BibitemShut {NoStop}%
\bibitem [{\citenamefont {Li}\ \emph {et~al.}(2018)\citenamefont {Li},
  \citenamefont {Tan},\ and\ \citenamefont {Jiang}}]{Li18}%
  \BibitemOpen
  \bibfield  {author} {\bibinfo {author} {\bibfnamefont {C.-D.}\ \bibnamefont
  {Li}}, \bibinfo {author} {\bibfnamefont {D.-R.}\ \bibnamefont {Tan}}, \ and\
  \bibinfo {author} {\bibfnamefont {F.-J.}\ \bibnamefont {Jiang}},\ }\bibfield
  {title} {\bibinfo {title} {\emph {Applications of neural networks to the
  studies of phase transitions of two-dimensional {Potts} models}},\ }\href
  {\doibase 10.1016/j.aop.2018.02.018} {\bibfield  {journal} {\bibinfo
  {journal} {Ann. Phys.}\ }\textbf {\bibinfo {volume} {391}},\ \bibinfo {pages}
  {312} (\bibinfo {year} {2018})}\BibitemShut {NoStop}%
\bibitem [{\citenamefont {Wang}(2016)}]{Wang16}%
  \BibitemOpen
  \bibfield  {author} {\bibinfo {author} {\bibfnamefont {L.}~\bibnamefont
  {Wang}},\ }\bibfield  {title} {\bibinfo {title} {\emph {Discovering phase
  transitions with unsupervised learning}},\ }\href {\doibase
  10.1103/PhysRevB.94.195105} {\bibfield  {journal} {\bibinfo  {journal} {Phys.
  Rev. B}\ }\textbf {\bibinfo {volume} {94}},\ \bibinfo {pages} {195105}
  (\bibinfo {year} {2016})}\BibitemShut {NoStop}%
\bibitem [{\citenamefont {Liu}\ and\ \citenamefont {van
  Nieuwenburg}(2018)}]{Liu18}%
  \BibitemOpen
  \bibfield  {author} {\bibinfo {author} {\bibfnamefont {Y.-H.}\ \bibnamefont
  {Liu}}\ and\ \bibinfo {author} {\bibfnamefont {E.~P.~L.}\ \bibnamefont {van
  Nieuwenburg}},\ }\bibfield  {title} {\bibinfo {title} {\emph {Discriminative
  Cooperative Networks for Detecting Phase Transitions}},\ }\href {\doibase
  10.1103/PhysRevLett.120.176401} {\bibfield  {journal} {\bibinfo  {journal}
  {Phys. Rev. Lett.}\ }\textbf {\bibinfo {volume} {120}},\ \bibinfo {pages}
  {176401} (\bibinfo {year} {2018})}\BibitemShut {NoStop}%
\bibitem [{\citenamefont {Broecker}\ \emph {et~al.}(2017)\citenamefont
  {Broecker}, \citenamefont {Carrasquilla}, \citenamefont {Melko},\ and\
  \citenamefont {Trebst}}]{Broecker17}%
  \BibitemOpen
  \bibfield  {author} {\bibinfo {author} {\bibfnamefont {P.}~\bibnamefont
  {Broecker}}, \bibinfo {author} {\bibfnamefont {J.}~\bibnamefont
  {Carrasquilla}}, \bibinfo {author} {\bibfnamefont {R.~G.}\ \bibnamefont
  {Melko}}, \ and\ \bibinfo {author} {\bibfnamefont {S.}~\bibnamefont
  {Trebst}},\ }\bibfield  {title} {\bibinfo {title} {\emph {Machine learning
  quantum phases of matter beyond the fermion sign problem}},\ }\href {\doibase
  10.1038/s41598-017-09098-0} {\bibfield  {journal} {\bibinfo  {journal} {Sci.
  Rep.}\ }\textbf {\bibinfo {volume} {7}},\ \bibinfo {pages} {8823} (\bibinfo
  {year} {2017})}\BibitemShut {NoStop}%
\bibitem [{\citenamefont {Huembeli}\ \emph {et~al.}(2019)\citenamefont
  {Huembeli}, \citenamefont {Dauphin}, \citenamefont {Wittek},\ and\
  \citenamefont {Gogolin}}]{Huembeli19}%
  \BibitemOpen
  \bibfield  {author} {\bibinfo {author} {\bibfnamefont {P.}~\bibnamefont
  {Huembeli}}, \bibinfo {author} {\bibfnamefont {A.}~\bibnamefont {Dauphin}},
  \bibinfo {author} {\bibfnamefont {P.}~\bibnamefont {Wittek}}, \ and\ \bibinfo
  {author} {\bibfnamefont {C.}~\bibnamefont {Gogolin}},\ }\bibfield  {title}
  {\bibinfo {title} {\emph {Automated discovery of characteristic features of
  phase transitions in many-body localization}},\ }\href {\doibase
  10.1103/PhysRevB.99.104106} {\bibfield  {journal} {\bibinfo  {journal} {Phys.
  Rev. B}\ }\textbf {\bibinfo {volume} {99}},\ \bibinfo {pages} {104106}
  (\bibinfo {year} {2019})}\BibitemShut {NoStop}%
\bibitem [{\citenamefont {Ch'ng}\ \emph {et~al.}(2018)\citenamefont {Ch'ng},
  \citenamefont {Vazquez},\ and\ \citenamefont {Khatami}}]{Chng18}%
  \BibitemOpen
  \bibfield  {author} {\bibinfo {author} {\bibfnamefont {K.}~\bibnamefont
  {Ch'ng}}, \bibinfo {author} {\bibfnamefont {N.}~\bibnamefont {Vazquez}}, \
  and\ \bibinfo {author} {\bibfnamefont {E.}~\bibnamefont {Khatami}},\
  }\bibfield  {title} {\bibinfo {title} {\emph {Unsupervised machine learning
  account of magnetic transitions in the Hubbard model}},\ }\href {\doibase
  10.1103/PhysRevE.97.013306} {\bibfield  {journal} {\bibinfo  {journal} {Phys.
  Rev. E}\ }\textbf {\bibinfo {volume} {97}},\ \bibinfo {pages} {013306}
  (\bibinfo {year} {2018})}\BibitemShut {NoStop}%
\bibitem [{\citenamefont {Th{\'e}veniaut}\ and\ \citenamefont
  {Alet}(2019)}]{Theveniaut19}%
  \BibitemOpen
  \bibfield  {author} {\bibinfo {author} {\bibfnamefont {H.}~\bibnamefont
  {Th{\'e}veniaut}}\ and\ \bibinfo {author} {\bibfnamefont {F.}~\bibnamefont
  {Alet}},\ }\bibfield  {title} {\bibinfo {title} {\emph {Neural network setups
  for a precise detection of the many-body localization transition: finite-size
  scaling and limitations}},\ }\href {\doibase 10.1103/PhysRevB.100.224202}
  {\bibfield  {journal} {\bibinfo  {journal} {Phys. Rev. B}\ }\textbf {\bibinfo
  {volume} {100}},\ \bibinfo {pages} {224202} (\bibinfo {year}
  {2019})}\BibitemShut {NoStop}%
\bibitem [{\citenamefont {Wetzel}(2017)}]{Wetzel17a}%
  \BibitemOpen
  \bibfield  {author} {\bibinfo {author} {\bibfnamefont {S.~J.}\ \bibnamefont
  {Wetzel}},\ }\bibfield  {title} {\bibinfo {title} {\emph {Unsupervised
  learning of phase transitions: From principal component analysis to
  variational autoencoders}},\ }\href {\doibase 10.1103/PhysRevE.96.022140}
  {\bibfield  {journal} {\bibinfo  {journal} {Phys. Rev. E}\ }\textbf {\bibinfo
  {volume} {96}},\ \bibinfo {pages} {022140} (\bibinfo {year}
  {2017})}\BibitemShut {NoStop}%
\bibitem [{\citenamefont {Kottmann}\ \emph {et~al.}(2020)\citenamefont
  {Kottmann}, \citenamefont {Huembeli}, \citenamefont {Lewenstein},\ and\
  \citenamefont {Acin}}]{kottmann2020unsupervised}%
  \BibitemOpen
  \bibfield  {author} {\bibinfo {author} {\bibfnamefont {K.}~\bibnamefont
  {Kottmann}}, \bibinfo {author} {\bibfnamefont {P.}~\bibnamefont {Huembeli}},
  \bibinfo {author} {\bibfnamefont {M.}~\bibnamefont {Lewenstein}}, \ and\
  \bibinfo {author} {\bibfnamefont {A.}~\bibnamefont {Acin}},\ }\bibfield
  {title} {\bibinfo {title} {\emph {Unsupervised phase discovery with deep
  anomaly detection}},\ }\href {https://arxiv.org/abs/2003.09905} {\bibfield
  {journal} {\bibinfo  {journal} {arXiv:2003.09905}\ } (\bibinfo {year}
  {2020})}\BibitemShut {NoStop}%
\bibitem [{\citenamefont {Vargas-Hern\'{a}ndez}\ \emph
  {et~al.}(2018)\citenamefont {Vargas-Hern\'{a}ndez}, \citenamefont {Sous},
  \citenamefont {Berciu},\ and\ \citenamefont {Krems}}]{Vargas18b}%
  \BibitemOpen
  \bibfield  {author} {\bibinfo {author} {\bibfnamefont {R.~A.}\ \bibnamefont
  {Vargas-Hern\'{a}ndez}}, \bibinfo {author} {\bibfnamefont {J.}~\bibnamefont
  {Sous}}, \bibinfo {author} {\bibfnamefont {M.}~\bibnamefont {Berciu}}, \ and\
  \bibinfo {author} {\bibfnamefont {R.~V.}\ \bibnamefont {Krems}},\ }\bibfield
  {title} {\bibinfo {title} {\emph {Extrapolating Quantum Observables with
  Machine Learning: Inferring Multiple Phase Transitions from Properties of a
  Single Phase}},\ }\href {\doibase 10.1103/PhysRevLett.121.255702} {\bibfield
  {journal} {\bibinfo  {journal} {Phys. Rev. Lett.}\ }\textbf {\bibinfo
  {volume} {121}},\ \bibinfo {pages} {255702} (\bibinfo {year}
  {2018})}\BibitemShut {NoStop}%
\bibitem [{\citenamefont {Deng}\ \emph {et~al.}(2017)\citenamefont {Deng},
  \citenamefont {Li},\ and\ \citenamefont {Das~Sarma}}]{Deng17}%
  \BibitemOpen
  \bibfield  {author} {\bibinfo {author} {\bibfnamefont {D.-L.}\ \bibnamefont
  {Deng}}, \bibinfo {author} {\bibfnamefont {X.}~\bibnamefont {Li}}, \ and\
  \bibinfo {author} {\bibfnamefont {S.}~\bibnamefont {Das~Sarma}},\ }\bibfield
  {title} {\bibinfo {title} {\emph {Quantum Entanglement in Neural Network
  States}},\ }\href {\doibase 10.1103/PhysRevX.7.021021} {\bibfield  {journal}
  {\bibinfo  {journal} {Phys. Rev. X}\ }\textbf {\bibinfo {volume} {7}},\
  \bibinfo {pages} {021021} (\bibinfo {year} {2017})}\BibitemShut {NoStop}%
\bibitem [{\citenamefont {Huembeli}\ \emph {et~al.}(2018)\citenamefont
  {Huembeli}, \citenamefont {Dauphin},\ and\ \citenamefont
  {Wittek}}]{Huembeli18}%
  \BibitemOpen
  \bibfield  {author} {\bibinfo {author} {\bibfnamefont {P.}~\bibnamefont
  {Huembeli}}, \bibinfo {author} {\bibfnamefont {A.}~\bibnamefont {Dauphin}}, \
  and\ \bibinfo {author} {\bibfnamefont {P.}~\bibnamefont {Wittek}},\
  }\bibfield  {title} {\bibinfo {title} {\emph {Identifying quantum phase
  transitions with adversarial neural networks}},\ }\href {\doibase
  10.1103/PhysRevB.97.134109} {\bibfield  {journal} {\bibinfo  {journal} {Phys.
  Rev. B}\ }\textbf {\bibinfo {volume} {97}},\ \bibinfo {pages} {134109}
  (\bibinfo {year} {2018})}\BibitemShut {NoStop}%
\bibitem [{\citenamefont {Zhang}\ \emph {et~al.}(2018)\citenamefont {Zhang},
  \citenamefont {Shen},\ and\ \citenamefont {Zhai}}]{Zhang18PRL}%
  \BibitemOpen
  \bibfield  {author} {\bibinfo {author} {\bibfnamefont {P.}~\bibnamefont
  {Zhang}}, \bibinfo {author} {\bibfnamefont {H.}~\bibnamefont {Shen}}, \ and\
  \bibinfo {author} {\bibfnamefont {H.}~\bibnamefont {Zhai}},\ }\bibfield
  {title} {\bibinfo {title} {\emph {Machine Learning Topological Invariants
  with Neural Networks}},\ }\href {\doibase 10.1103/PhysRevLett.120.066401}
  {\bibfield  {journal} {\bibinfo  {journal} {Phys. Rev. Lett.}\ }\textbf
  {\bibinfo {volume} {120}},\ \bibinfo {pages} {066401} (\bibinfo {year}
  {2018})}\BibitemShut {NoStop}%
\bibitem [{\citenamefont {Tsai}\ \emph {et~al.}(2020)\citenamefont {Tsai},
  \citenamefont {Yu}, \citenamefont {Hsu},\ and\ \citenamefont
  {Chung}}]{Tsai19}%
  \BibitemOpen
  \bibfield  {author} {\bibinfo {author} {\bibfnamefont {Y.-H.}\ \bibnamefont
  {Tsai}}, \bibinfo {author} {\bibfnamefont {M.-Z.}\ \bibnamefont {Yu}},
  \bibinfo {author} {\bibfnamefont {Y.-H.}\ \bibnamefont {Hsu}}, \ and\
  \bibinfo {author} {\bibfnamefont {M.-C.}\ \bibnamefont {Chung}},\ }\bibfield
  {title} {\bibinfo {title} {\emph {Deep learning of topological phase
  transitions from entanglement aspects}},\ }\href {\doibase
  10.1103/PhysRevB.102.054512} {\bibfield  {journal} {\bibinfo  {journal}
  {Phys. Rev. B}\ }\textbf {\bibinfo {volume} {102}},\ \bibinfo {pages}
  {054512} (\bibinfo {year} {2020})}\BibitemShut {NoStop}%
\bibitem [{\citenamefont {Greplova}\ \emph {et~al.}(2020)\citenamefont
  {Greplova}, \citenamefont {Valenti}, \citenamefont {Boschung}, \citenamefont
  {Sch{\"a}fer}, \citenamefont {L{\"o}rch},\ and\ \citenamefont
  {Huber}}]{Greplova19}%
  \BibitemOpen
  \bibfield  {author} {\bibinfo {author} {\bibfnamefont {E.}~\bibnamefont
  {Greplova}}, \bibinfo {author} {\bibfnamefont {A.}~\bibnamefont {Valenti}},
  \bibinfo {author} {\bibfnamefont {G.}~\bibnamefont {Boschung}}, \bibinfo
  {author} {\bibfnamefont {F.}~\bibnamefont {Sch{\"a}fer}}, \bibinfo {author}
  {\bibfnamefont {N.}~\bibnamefont {L{\"o}rch}}, \ and\ \bibinfo {author}
  {\bibfnamefont {S.}~\bibnamefont {Huber}},\ }\bibfield  {title} {\bibinfo
  {title} {\emph {Unsupervised identification of topological phase transitions
  using predictive models}},\ }\href {\doibase 10.1088/1367-2630/ab7771}
  {\bibfield  {journal} {\bibinfo  {journal} {New J. Phys.}\ }\textbf {\bibinfo
  {volume} {22}},\ \bibinfo {pages} {045003} (\bibinfo {year}
  {2020})}\BibitemShut {NoStop}%
\bibitem [{\citenamefont {Rem}\ \emph {et~al.}(2019)\citenamefont {Rem},
  \citenamefont {K\"{a}ming}, \citenamefont {Tarnowski}, \citenamefont
  {Asteria}, \citenamefont {Fl\"{a}schner}, \citenamefont {Becker},
  \citenamefont {Sengstock},\ and\ \citenamefont {Weitenberg}}]{Rem19}%
  \BibitemOpen
  \bibfield  {author} {\bibinfo {author} {\bibfnamefont {B.~S.}\ \bibnamefont
  {Rem}}, \bibinfo {author} {\bibfnamefont {N.}~\bibnamefont {K\"{a}ming}},
  \bibinfo {author} {\bibfnamefont {M.}~\bibnamefont {Tarnowski}}, \bibinfo
  {author} {\bibfnamefont {L.}~\bibnamefont {Asteria}}, \bibinfo {author}
  {\bibfnamefont {N.}~\bibnamefont {Fl\"{a}schner}}, \bibinfo {author}
  {\bibfnamefont {C.}~\bibnamefont {Becker}}, \bibinfo {author} {\bibfnamefont
  {K.}~\bibnamefont {Sengstock}}, \ and\ \bibinfo {author} {\bibfnamefont
  {C.}~\bibnamefont {Weitenberg}},\ }\bibfield  {title} {\bibinfo {title}
  {\emph {Identifying quantum phase transitions using artificial neural
  networks on experimental data}},\ }\href {\doibase 10.1038/s41567-019-0554-0}
  {\bibfield  {journal} {\bibinfo  {journal} {Nat. Phys.}\ }\textbf {\bibinfo
  {volume} {15}},\ \bibinfo {pages} {917–920} (\bibinfo {year}
  {2019})}\BibitemShut {NoStop}%
\bibitem [{\citenamefont {Khatami}\ \emph {et~al.}(2020)\citenamefont
  {Khatami}, \citenamefont {Guardado-Sanchez}, \citenamefont {Spar},
  \citenamefont {Carrasquilla}, \citenamefont {Bakr},\ and\ \citenamefont
  {Scalettar}}]{Khatami20}%
  \BibitemOpen
  \bibfield  {author} {\bibinfo {author} {\bibfnamefont {E.}~\bibnamefont
  {Khatami}}, \bibinfo {author} {\bibfnamefont {E.}~\bibnamefont
  {Guardado-Sanchez}}, \bibinfo {author} {\bibfnamefont {B.~M.}\ \bibnamefont
  {Spar}}, \bibinfo {author} {\bibfnamefont {J.~F.}\ \bibnamefont
  {Carrasquilla}}, \bibinfo {author} {\bibfnamefont {W.~S.}\ \bibnamefont
  {Bakr}}, \ and\ \bibinfo {author} {\bibfnamefont {R.~T.}\ \bibnamefont
  {Scalettar}},\ }\bibfield  {title} {\bibinfo {title} {\emph {Visualizing
  strange metallic correlations in the two-dimensional Fermi-Hubbard model with
  artificial intelligence}},\ }\href {\doibase 10.1103/PhysRevA.102.033326}
  {\bibfield  {journal} {\bibinfo  {journal} {Phys. Rev. A}\ }\textbf {\bibinfo
  {volume} {102}},\ \bibinfo {pages} {033326} (\bibinfo {year}
  {2020})}\BibitemShut {NoStop}%
\bibitem [{\citenamefont {Zhang}\ and\ \citenamefont {Kim}(2017)}]{Kim17}%
  \BibitemOpen
  \bibfield  {author} {\bibinfo {author} {\bibfnamefont {Y.}~\bibnamefont
  {Zhang}}\ and\ \bibinfo {author} {\bibfnamefont {E.-A.}\ \bibnamefont
  {Kim}},\ }\bibfield  {title} {\bibinfo {title} {\emph {Quantum Loop
  Topography for Machine Learning}},\ }\href {\doibase
  10.1103/PhysRevLett.118.216401} {\bibfield  {journal} {\bibinfo  {journal}
  {Phys. Rev. Lett.}\ }\textbf {\bibinfo {volume} {118}},\ \bibinfo {pages}
  {216401} (\bibinfo {year} {2017})}\BibitemShut {NoStop}%
\bibitem [{\citenamefont {Beach}\ \emph {et~al.}(2018)\citenamefont {Beach},
  \citenamefont {Golubeva},\ and\ \citenamefont {Melko}}]{Beach18}%
  \BibitemOpen
  \bibfield  {author} {\bibinfo {author} {\bibfnamefont {M.~J.~S.}\
  \bibnamefont {Beach}}, \bibinfo {author} {\bibfnamefont {A.}~\bibnamefont
  {Golubeva}}, \ and\ \bibinfo {author} {\bibfnamefont {R.~G.}\ \bibnamefont
  {Melko}},\ }\bibfield  {title} {\bibinfo {title} {\emph {Machine learning
  vortices at the {Kosterlitz-Thouless} transition}},\ }\href {\doibase
  10.1103/PhysRevB.97.045207} {\bibfield  {journal} {\bibinfo  {journal} {Phys.
  Rev. B}\ }\textbf {\bibinfo {volume} {97}},\ \bibinfo {pages} {045207}
  (\bibinfo {year} {2018})}\BibitemShut {NoStop}%
\bibitem [{\citenamefont {Richter-Laskowska}\ \emph {et~al.}(2018)\citenamefont
  {Richter-Laskowska}, \citenamefont {Khan}, \citenamefont {Trivedi},\ and\
  \citenamefont {Ma\'{s}ka}}]{Richter18}%
  \BibitemOpen
  \bibfield  {author} {\bibinfo {author} {\bibfnamefont {M.}~\bibnamefont
  {Richter-Laskowska}}, \bibinfo {author} {\bibfnamefont {H.}~\bibnamefont
  {Khan}}, \bibinfo {author} {\bibfnamefont {N.}~\bibnamefont {Trivedi}}, \
  and\ \bibinfo {author} {\bibfnamefont {M.~M.}\ \bibnamefont {Ma\'{s}ka}},\
  }\bibfield  {title} {\bibinfo {title} {\emph {A machine learning approach to
  the Berezinskii-Kosterlitz-Thouless transition in classical and quantum
  models}},\ }\href {\doibase 10.5488/CMP.21.33602} {\bibfield  {journal}
  {\bibinfo  {journal} {Condens. Matter Phys.}\ }\textbf {\bibinfo {volume}
  {21}},\ \bibinfo {pages} {33602} (\bibinfo {year} {2018})}\BibitemShut
  {NoStop}%
\bibitem [{\citenamefont {Guidotti}\ \emph {et~al.}(2018)\citenamefont
  {Guidotti}, \citenamefont {Monreale}, \citenamefont {Ruggieri}, \citenamefont
  {Turini}, \citenamefont {Pedreschi},\ and\ \citenamefont
  {Giannotti}}]{Guidotti18}%
  \BibitemOpen
  \bibfield  {author} {\bibinfo {author} {\bibfnamefont {R.}~\bibnamefont
  {Guidotti}}, \bibinfo {author} {\bibfnamefont {A.}~\bibnamefont {Monreale}},
  \bibinfo {author} {\bibfnamefont {S.}~\bibnamefont {Ruggieri}}, \bibinfo
  {author} {\bibfnamefont {F.}~\bibnamefont {Turini}}, \bibinfo {author}
  {\bibfnamefont {D.}~\bibnamefont {Pedreschi}}, \ and\ \bibinfo {author}
  {\bibfnamefont {F.}~\bibnamefont {Giannotti}},\ }\bibfield  {title} {\bibinfo
  {title} {\emph {A Survey Of Methods For Explaining Black Box Models}},\
  }\href {https://arxiv.org/abs/1802.01933} {\bibfield  {journal} {\bibinfo
  {journal} {arXiv:1802.01933}\ } (\bibinfo {year} {2018})}\BibitemShut
  {NoStop}%
\bibitem [{\citenamefont {Doshi-Velez}\ and\ \citenamefont
  {Kim}(2017)}]{DoshiVelez17}%
  \BibitemOpen
  \bibfield  {author} {\bibinfo {author} {\bibfnamefont {F.}~\bibnamefont
  {Doshi-Velez}}\ and\ \bibinfo {author} {\bibfnamefont {B.}~\bibnamefont
  {Kim}},\ }\bibfield  {title} {\bibinfo {title} {\emph {Towards A Rigorous
  Science of Interpretable Machine Learning}},\ }\href
  {https://arxiv.org/abs/1702.08608} {\bibfield  {journal} {\bibinfo  {journal}
  {arXiv:1702.08608}\ } (\bibinfo {year} {2017})}\BibitemShut {NoStop}%
\bibitem [{\citenamefont {Ponte}\ and\ \citenamefont {Melko}(2017)}]{Ponte17}%
  \BibitemOpen
  \bibfield  {author} {\bibinfo {author} {\bibfnamefont {P.}~\bibnamefont
  {Ponte}}\ and\ \bibinfo {author} {\bibfnamefont {R.~G.}\ \bibnamefont
  {Melko}},\ }\bibfield  {title} {\bibinfo {title} {\emph {Kernel methods for
  interpretable machine learning of order parameters}},\ }\href {\doibase
  10.1103/PhysRevB.96.205146} {\bibfield  {journal} {\bibinfo  {journal} {Phys.
  Rev. B}\ }\textbf {\bibinfo {volume} {96}},\ \bibinfo {pages} {205146}
  (\bibinfo {year} {2017})}\BibitemShut {NoStop}%
\bibitem [{\citenamefont {Zhang}\ \emph {et~al.}(2019)\citenamefont {Zhang},
  \citenamefont {Wang},\ and\ \citenamefont {Wang}}]{Zhang19a}%
  \BibitemOpen
  \bibfield  {author} {\bibinfo {author} {\bibfnamefont {W.}~\bibnamefont
  {Zhang}}, \bibinfo {author} {\bibfnamefont {L.}~\bibnamefont {Wang}}, \ and\
  \bibinfo {author} {\bibfnamefont {Z.}~\bibnamefont {Wang}},\ }\bibfield
  {title} {\bibinfo {title} {\emph {Interpretable machine learning study of the
  many-body localization transition in disordered quantum {Ising} spin
  chains}},\ }\href {\doibase 10.1103/PhysRevB.99.054208} {\bibfield  {journal}
  {\bibinfo  {journal} {Phys. Rev. B}\ }\textbf {\bibinfo {volume} {99}},\
  \bibinfo {pages} {054208} (\bibinfo {year} {2019})}\BibitemShut {NoStop}%
\bibitem [{\citenamefont {Greitemann}\ \emph {et~al.}(2019)\citenamefont
  {Greitemann}, \citenamefont {Liu}, \citenamefont {Jaubert}, \citenamefont
  {Yan}, \citenamefont {Shannon},\ and\ \citenamefont {Pollet}}]{Greitemann19}%
  \BibitemOpen
  \bibfield  {author} {\bibinfo {author} {\bibfnamefont {J.}~\bibnamefont
  {Greitemann}}, \bibinfo {author} {\bibfnamefont {K.}~\bibnamefont {Liu}},
  \bibinfo {author} {\bibfnamefont {L.~D.~C.}\ \bibnamefont {Jaubert}},
  \bibinfo {author} {\bibfnamefont {H.}~\bibnamefont {Yan}}, \bibinfo {author}
  {\bibfnamefont {N.}~\bibnamefont {Shannon}}, \ and\ \bibinfo {author}
  {\bibfnamefont {L.}~\bibnamefont {Pollet}},\ }\bibfield  {title} {\bibinfo
  {title} {\emph {Identification of emergent constraints and hidden order in
  frustrated magnets using tensorial kernel methods of machine learning}},\
  }\href {\doibase 10.1103/PhysRevB.100.174408} {\bibfield  {journal} {\bibinfo
   {journal} {Phys. Rev. B}\ }\textbf {\bibinfo {volume} {100}} (\bibinfo
  {year} {2019}),\ 10.1103/PhysRevB.100.174408}\BibitemShut {NoStop}%
\bibitem [{\citenamefont {Greitemann}\ \emph {et~al.}(2020)\citenamefont
  {Greitemann}, \citenamefont {Liu},\ and\ \citenamefont
  {Pollet}}]{Greitemann20}%
  \BibitemOpen
  \bibfield  {author} {\bibinfo {author} {\bibfnamefont {J.}~\bibnamefont
  {Greitemann}}, \bibinfo {author} {\bibfnamefont {K.}~\bibnamefont {Liu}}, \
  and\ \bibinfo {author} {\bibfnamefont {L.}~\bibnamefont {Pollet}},\
  }\bibfield  {title} {\bibinfo {title} {\emph {The view of TK-SVM on the phase
  hierarchy in the classical kagome Heisenberg antiferromagnet}},\ }\href
  {https://arxiv.org/abs/2007.01685} {\bibfield  {journal} {\bibinfo  {journal}
  {arXiv:2007.01685}\ } (\bibinfo {year} {2020})}\BibitemShut {NoStop}%
\bibitem [{\citenamefont {Wetzel}\ and\ \citenamefont
  {Scherzer}(2017)}]{Wetzel17b}%
  \BibitemOpen
  \bibfield  {author} {\bibinfo {author} {\bibfnamefont {S.~J.}\ \bibnamefont
  {Wetzel}}\ and\ \bibinfo {author} {\bibfnamefont {M.}~\bibnamefont
  {Scherzer}},\ }\bibfield  {title} {\bibinfo {title} {\emph {Machine learning
  of explicit order parameters: From the Ising model to SU(2) lattice gauge
  theory}},\ }\href {\doibase 10.1103/PhysRevB.96.184410} {\bibfield  {journal}
  {\bibinfo  {journal} {Phys. Rev. B}\ }\textbf {\bibinfo {volume} {96}},\
  \bibinfo {pages} {184410} (\bibinfo {year} {2017})}\BibitemShut {NoStop}%
\bibitem [{\citenamefont {Wetzel}\ \emph {et~al.}(2020)\citenamefont {Wetzel},
  \citenamefont {Melko}, \citenamefont {Scott}, \citenamefont {Panju},\ and\
  \citenamefont {Ganesh}}]{Wetzel20}%
  \BibitemOpen
  \bibfield  {author} {\bibinfo {author} {\bibfnamefont {S.~J.}\ \bibnamefont
  {Wetzel}}, \bibinfo {author} {\bibfnamefont {R.~G.}\ \bibnamefont {Melko}},
  \bibinfo {author} {\bibfnamefont {J.}~\bibnamefont {Scott}}, \bibinfo
  {author} {\bibfnamefont {M.}~\bibnamefont {Panju}}, \ and\ \bibinfo {author}
  {\bibfnamefont {V.}~\bibnamefont {Ganesh}},\ }\bibfield  {title} {\bibinfo
  {title} {\emph {Discovering symmetry invariants and conserved quantities by
  interpreting siamese neural networks}},\ }\href {\doibase
  10.1103/PhysRevResearch.2.033499} {\bibfield  {journal} {\bibinfo  {journal}
  {Phys. Rev. Research}\ }\textbf {\bibinfo {volume} {2}},\ \bibinfo {pages}
  {033499} (\bibinfo {year} {2020})}\BibitemShut {NoStop}%
\bibitem [{\citenamefont {Zhang}\ \emph {et~al.}(2020)\citenamefont {Zhang},
  \citenamefont {Ginsparg},\ and\ \citenamefont {Kim}}]{Zhang19b}%
  \BibitemOpen
  \bibfield  {author} {\bibinfo {author} {\bibfnamefont {Y.}~\bibnamefont
  {Zhang}}, \bibinfo {author} {\bibfnamefont {P.}~\bibnamefont {Ginsparg}}, \
  and\ \bibinfo {author} {\bibfnamefont {E.-A.}\ \bibnamefont {Kim}},\
  }\bibfield  {title} {\bibinfo {title} {\emph {Interpreting machine learning
  of topological quantum phase transitions}},\ }\href {\doibase
  10.1103/PhysRevResearch.2.023283} {\bibfield  {journal} {\bibinfo  {journal}
  {Phys. Rev. Research}\ }\textbf {\bibinfo {volume} {2}},\ \bibinfo {pages}
  {023283} (\bibinfo {year} {2020})}\BibitemShut {NoStop}%
\bibitem [{\citenamefont {Cook}(1977)}]{Cook77}%
  \BibitemOpen
  \bibfield  {author} {\bibinfo {author} {\bibfnamefont {R.~D.}\ \bibnamefont
  {Cook}},\ }\bibfield  {title} {\bibinfo {title} {\emph {Detection of
  Influential Observation in Linear Regression}},\ }\href {\doibase
  10.2307/1268249} {\bibfield  {journal} {\bibinfo  {journal} {Technometrics}\
  }\textbf {\bibinfo {volume} {19}},\ \bibinfo {pages} {15} (\bibinfo {year}
  {1977})}\BibitemShut {NoStop}%
\bibitem [{\citenamefont {Cook}\ and\ \citenamefont {Weisberg}(1980)}]{Cook80}%
  \BibitemOpen
  \bibfield  {author} {\bibinfo {author} {\bibfnamefont {R.~D.}\ \bibnamefont
  {Cook}}\ and\ \bibinfo {author} {\bibfnamefont {S.}~\bibnamefont
  {Weisberg}},\ }\bibfield  {title} {\bibinfo {title} {\emph {Characterizations
  of an Empirical Influence Function for Detecting Influential Cases in
  Regression}},\ }\href {\doibase 10.2307/1268187} {\bibfield  {journal}
  {\bibinfo  {journal} {Technometrics}\ }\textbf {\bibinfo {volume} {22}},\
  \bibinfo {pages} {495} (\bibinfo {year} {1980})}\BibitemShut {NoStop}%
\bibitem [{\citenamefont {Cook}\ and\ \citenamefont {Weisberg}(1982)}]{Cook82}%
  \BibitemOpen
  \bibfield  {author} {\bibinfo {author} {\bibfnamefont {R.~D.}\ \bibnamefont
  {Cook}}\ and\ \bibinfo {author} {\bibfnamefont {S.}~\bibnamefont
  {Weisberg}},\ }\href@noop {} {\emph {\bibinfo {title} {Residuals and
  Influence in Regression}}}\ (\bibinfo  {publisher} {Chapman and Hall},\
  \bibinfo {address} {New York and London},\ \bibinfo {year}
  {1982})\BibitemShut {NoStop}%
\bibitem [{\citenamefont {Sagun}\ \emph {et~al.}(2017)\citenamefont {Sagun},
  \citenamefont {Bottou},\ and\ \citenamefont {LeCun}}]{Sagun17}%
  \BibitemOpen
  \bibfield  {author} {\bibinfo {author} {\bibfnamefont {L.}~\bibnamefont
  {Sagun}}, \bibinfo {author} {\bibfnamefont {L.}~\bibnamefont {Bottou}}, \
  and\ \bibinfo {author} {\bibfnamefont {Y.}~\bibnamefont {LeCun}},\ }\bibfield
   {title} {\bibinfo {title} {\emph {Eigenvalues of the Hessian in Deep
  Learning: Singularity and Beyond}},\ }\href
  {https://arxiv.org/abs/1611.07476} {\bibfield  {journal} {\bibinfo  {journal}
  {arXiv:1611.07476v2}\ } (\bibinfo {year} {2017})}\BibitemShut {NoStop}%
\bibitem [{\citenamefont {Koh}\ and\ \citenamefont {Liang}(2017)}]{Koh17}%
  \BibitemOpen
  \bibfield  {author} {\bibinfo {author} {\bibfnamefont {P.~W.}\ \bibnamefont
  {Koh}}\ and\ \bibinfo {author} {\bibfnamefont {P.}~\bibnamefont {Liang}},\
  }\bibfield  {title} {\bibinfo {title} {\emph {Understanding Black-box
  Predictions via Influence Functions}},\ }in\ \href
  {http://proceedings.mlr.press/v70/koh17a.html} {\emph {\bibinfo {booktitle}
  {{Proceedings of the 34th International Conference} on {Machine Learning}}}}\
  (\bibinfo {year} {2017})\BibitemShut {NoStop}%
\bibitem [{\citenamefont {Koh}\ \emph {et~al.}(2019)\citenamefont {Koh},
  \citenamefont {Ang}, \citenamefont {Teo},\ and\ \citenamefont
  {Liang}}]{Koh19}%
  \BibitemOpen
  \bibfield  {author} {\bibinfo {author} {\bibfnamefont {P.~W.}\ \bibnamefont
  {Koh}}, \bibinfo {author} {\bibfnamefont {K.-S.}\ \bibnamefont {Ang}},
  \bibinfo {author} {\bibfnamefont {H.~H.~K.}\ \bibnamefont {Teo}}, \ and\
  \bibinfo {author} {\bibfnamefont {P.}~\bibnamefont {Liang}},\ }\bibfield
  {title} {\bibinfo {title} {\emph {On the Accuracy of Influence Functions for
  Measuring Group Effects}},\ }\href {https://arxiv.org/abs/1905.13289}
  {\bibfield  {journal} {\bibinfo  {journal} {arXiv:1905.13289}\ } (\bibinfo
  {year} {2019})}\BibitemShut {NoStop}%
\bibitem [{\citenamefont {Dawid}\ \emph {et~al.}(2020)\citenamefont {Dawid},
  \citenamefont {Huembeli}, \citenamefont {Tomza}, \citenamefont {Lewenstein},\
  and\ \citenamefont {Dauphin}}]{OurRepo}%
  \BibitemOpen
  \bibfield  {author} {\bibinfo {author} {\bibfnamefont {A.}~\bibnamefont
  {Dawid}}, \bibinfo {author} {\bibfnamefont {P.}~\bibnamefont {Huembeli}},
  \bibinfo {author} {\bibfnamefont {M.}~\bibnamefont {Tomza}}, \bibinfo
  {author} {\bibfnamefont {M.}~\bibnamefont {Lewenstein}}, \ and\ \bibinfo
  {author} {\bibfnamefont {A.}~\bibnamefont {Dauphin}},\ }\href {\doibase
  10.5281/zenodo.3759432} {}\bibinfo {howpublished}
  {\url{http://doi.org/10.5281/zenodo.3759432}} (\bibinfo {year} {2020}),\
  \bibinfo {note} {{GitHub repository: Interpretable-Phase-Classification
  (Version arXiv1.1).}}\BibitemShut {Stop}%
\bibitem [{\citenamefont {Dutta}\ \emph {et~al.}(2015)\citenamefont {Dutta},
  \citenamefont {Gajda}, \citenamefont {Hauke}, \citenamefont {Lewenstein},
  \citenamefont {L\"{u}hmann}, \citenamefont {Malomed}, \citenamefont
  {Sowi\'{n}ski},\ and\ \citenamefont {Zakrzewski}}]{Dutta15}%
  \BibitemOpen
  \bibfield  {author} {\bibinfo {author} {\bibfnamefont {O.}~\bibnamefont
  {Dutta}}, \bibinfo {author} {\bibfnamefont {M.}~\bibnamefont {Gajda}},
  \bibinfo {author} {\bibfnamefont {P.}~\bibnamefont {Hauke}}, \bibinfo
  {author} {\bibfnamefont {M.}~\bibnamefont {Lewenstein}}, \bibinfo {author}
  {\bibfnamefont {D.-S.}\ \bibnamefont {L\"{u}hmann}}, \bibinfo {author}
  {\bibfnamefont {B.~A.}\ \bibnamefont {Malomed}}, \bibinfo {author}
  {\bibfnamefont {T.}~\bibnamefont {Sowi\'{n}ski}}, \ and\ \bibinfo {author}
  {\bibfnamefont {J.}~\bibnamefont {Zakrzewski}},\ }\bibfield  {title}
  {\bibinfo {title} {\emph {Non-standard {Hubbard} models in optical lattices:
  a review}},\ }\href@noop {} {\bibfield  {journal} {\bibinfo  {journal} {Rep.
  Prog. Phys.}\ }\textbf {\bibinfo {volume} {78}},\ \bibinfo {pages} {066001}
  (\bibinfo {year} {2015})}\BibitemShut {NoStop}%
\bibitem [{\citenamefont {Hallberg}\ \emph {et~al.}(1990)\citenamefont
  {Hallberg}, \citenamefont {Gagliano},\ and\ \citenamefont
  {Balseiro}}]{Hallberg90}%
  \BibitemOpen
  \bibfield  {author} {\bibinfo {author} {\bibfnamefont {E.}~\bibnamefont
  {Hallberg}}, \bibinfo {author} {\bibfnamefont {E.}~\bibnamefont {Gagliano}},
  \ and\ \bibinfo {author} {\bibfnamefont {C.}~\bibnamefont {Balseiro}},\
  }\bibfield  {title} {\bibinfo {title} {\emph {Finite-size study of a spin-1/2
  Heisenberg chain with competing interactions: Phase diagram and critical
  behavior}},\ }\href {\doibase 10.1103/PhysRevB.41.9474} {\bibfield  {journal}
  {\bibinfo  {journal} {Phys. Rev. B}\ }\textbf {\bibinfo {volume} {41}},\
  \bibinfo {pages} {9474} (\bibinfo {year} {1990})}\BibitemShut {NoStop}%
\bibitem [{\citenamefont {Mishra}\ \emph {et~al.}(2011)\citenamefont {Mishra},
  \citenamefont {Carrasquilla},\ and\ \citenamefont {Rigol}}]{Mishra11}%
  \BibitemOpen
  \bibfield  {author} {\bibinfo {author} {\bibfnamefont {T.}~\bibnamefont
  {Mishra}}, \bibinfo {author} {\bibfnamefont {J.}~\bibnamefont
  {Carrasquilla}}, \ and\ \bibinfo {author} {\bibfnamefont {M.}~\bibnamefont
  {Rigol}},\ }\bibfield  {title} {\bibinfo {title} {\emph {Phase diagram of the
  half-filled one-dimensional t-V-V' model}},\ }\href {\doibase
  10.1103/PhysRevB.84.115135} {\bibfield  {journal} {\bibinfo  {journal} {Phys.
  Rev. B}\ }\textbf {\bibinfo {volume} {84}},\ \bibinfo {pages} {115135}
  (\bibinfo {year} {2011})}\BibitemShut {NoStop}%
\bibitem [{\citenamefont {Weinberg}\ and\ \citenamefont
  {Bukov}(2017)}]{Weinberg17}%
  \BibitemOpen
  \bibfield  {author} {\bibinfo {author} {\bibfnamefont {P.}~\bibnamefont
  {Weinberg}}\ and\ \bibinfo {author} {\bibfnamefont {M.}~\bibnamefont
  {Bukov}},\ }\bibfield  {title} {\bibinfo {title} {\emph {QuSpin: a Python
  package for dynamics and exact diagonalisation of quantum many body systems
  part I: spin chains}},\ }\href {\doibase 10.21468/SciPostPhys.2.1.003}
  {\bibfield  {journal} {\bibinfo  {journal} {SciPost Phys.}\ }\textbf
  {\bibinfo {volume} {2}},\ \bibinfo {pages} {003} (\bibinfo {year}
  {2017})}\BibitemShut {NoStop}%
\bibitem [{\citenamefont {{P. Virtanen \textit{et al.}}}(2020)}]{SciPy}%
  \BibitemOpen
  \bibfield  {author} {\bibinfo {author} {\bibnamefont {{P. Virtanen \textit{et
  al.}}}},\ }\bibfield  {title} {\bibinfo {title} {\emph {{SciPy 1.0:
  Fundamental Algorithms for Scientific Computing in Python}}},\ }\href
  {\doibase https://doi.org/10.1038/s41592-019-0686-2} {\bibfield  {journal}
  {\bibinfo  {journal} {Nat. Methods}\ }\textbf {\bibinfo {volume} {17}},\
  \bibinfo {pages} {261} (\bibinfo {year} {2020})}\BibitemShut {NoStop}%
\end{thebibliography}
%

\newpage
\onecolumngrid

\appendix

\section{Phase diagram of the extended one-dimensional half-filled spinless Fermi-Hubbard model}
\label{app:PD}


We study the one-dimensional system consisting of spinless fermions at half-filling with hopping between neighboring sites with amplitudes $J$, interacting with nearest neighbors with strength $V_1$ and next-nearest neighbors with strength $V_2$, described with Hamiltonian \ref{eq:ham}.
The model exhibits four different phases, two of them co-exist in the limited range of parameters.
Without the next-nearest-neighbor interaction, $V_2$, the system can follow only patterns of the gapless liquid Luttinger (algebraic) phase (LL) or the charge-density wave of type I (CDW-I) with the degenerated density pattern 101010.
The CDW-I order parameter describing this transition reads $O_{\text{CDW-I}}=\frac{1}{L} \sum_{\la i,j \ra} |n_{i} - n_{j}|$, where $\la \ra$ symbolizes nearest neighbors.
The next-nearest-neighbor interaction, $V_2$ competes with $V_1$, so for non-zero $V_2$ but still smaller than $V_1$ the transition between LL and CDW-I shifts towards bigger $V_1$.
For sufficiently strong $V_2$ the bond-order (BO) phase emerges with the order parameter $O_{\text{BO}}=\frac{1}{L} \sum_{i}(-1)^{i} B_{i}$, where $B_{i}=\left\la c_{i}^{\dagger} c_{i+1}+c_{i+1}^{\dagger} c_{i}\right\ra$.
It turns into the charge-density wave of the type II (CDW-II) with the degenerated density pattern 11001100 for large $V_2$ values, with $O_{\text{CDW-II}}=\frac{1}{L} \sum_{\la \la i,j \ra \ra} |n_{i} - n_{j}|$, where $\la \la \ra \ra$ symbolizes next-nearest neighbors.

To calculate the ground states and order parameters of the model, we use QuSpin package~\cite{Weinberg17} to write the Hamiltonian for a 12-site system in the Fock basis, resulting in 924 basis states.
We assume periodic boundary conditions.
We perform the exact diagonalization with the SciPy package~\cite{SciPy}.
The ground states belonging to BO, CDW-I, and II phases are degenerated.
To lift the degeneracy of the ground state, we apply symmetry breaking (guiding) fields favoring one of the patterns.

This approach results in the order parameters in the LL being not exactly constant and equal to zero.
Instead, their values are growing very slowly when approaching the transition points.
Therefore, there is no exact transition point, so we define it as such parameters of the system that correspond to the order parameter being ten times bigger than the corresponding symmetry breaking fields.
Due to the guiding fields of values $10^{-7}$, $10^{-5}$, and $10^{-4}$ for 101010 and 11001100 density patterns and 1010 hopping pattern, respectively, the order parameters of values $10^{-6}$, $10^{-4}$, and $10^{-3}$ signal the transition to the CDW-I, CDW-II, and BO phase, respectively.

\begin{figure}[b]
\begin{center}
\includegraphics[width=\columnwidth]{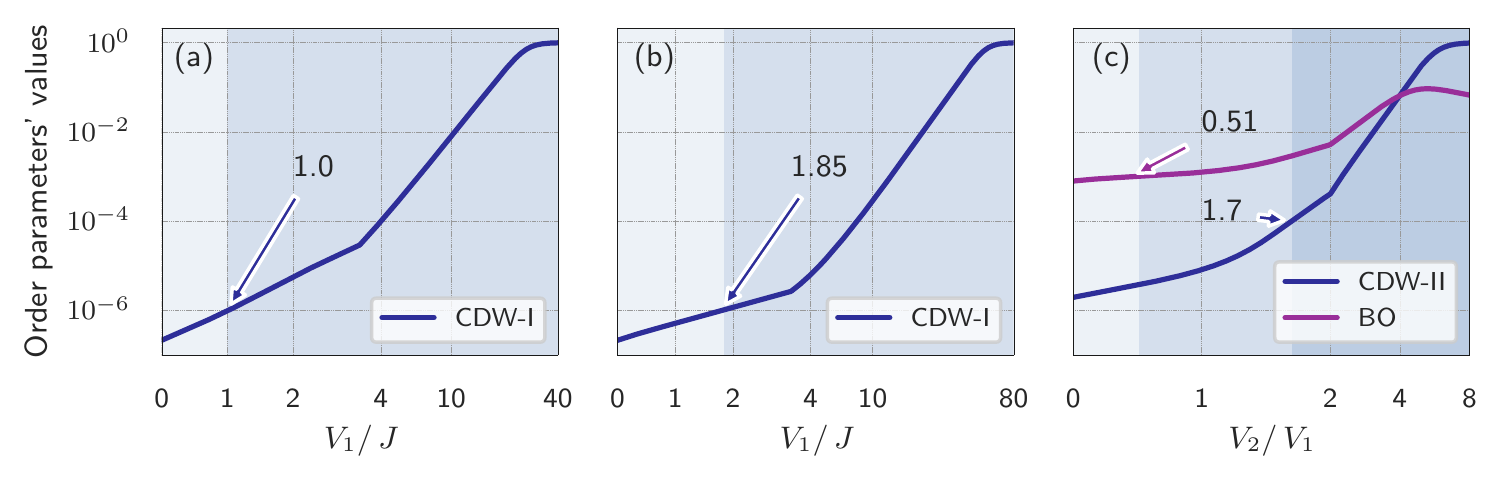}
\end{center}\vspace{-0.6cm}
\caption{Corresponding order parameters' values for three transition lines studied within this work, indicated with arrows (1)-(3) in Fig.~\ref{fig:intro}(b). (a)-(b) CDW-I order parameter for the transition line between the LL and the CDW-I phase for $V_2=0$ and $0.25 V_1$, respectively. (c) CDW-II and BO order parameters for the transition line between LL, BO, and CDW-II for $V_1=1/J$. Note the logarithmic scale of $y$-axis, and the symmetric log scale of $x$-axis with threshold points chosen to be 3, 3, and 2, respectively. Cusps in the lines are artificial and result from the symmetric log scale of $x$-axis.}
\label{fig:OP}
\end{figure}

The non-zero order parameter in the uniform phase and numerical arbitrariness of choosing the transition points are the main reasons why the influence functions' values in the LL phase, seen in Figs. 2, 3, and panels (a)-(b) of Fig. 4 of the manuscript are not precisely the same.
The third reason is the finite-size effects.
As the order parameters in the LL phase are growing very slowly, finally, the most helpful points are the ones near the transition - they are also the most unique from the training points labeled as LL, and the information they provide is the most valuable.
In the perfect scenario (observed, for example, for training on states obtained from mean-field calculations), the five most influential points randomly distribute over the whole LL phase.

It is interesting to note that the results presented in this work stay the same without the symmetry-breaking fields and do not depend on the size of the system.

Within this work, we train the convolutional neural network on three transition lines indicated with arrows (1)-(3) in Fig.~\ref{fig:intro}(b).
The first transition line leads from the LL to the CDW-I phase.
We calculate it for a constant $V_2=0$ and $V_1/J = \la 0, 40 \ra$.
It is a source of training data for both Figs.~\ref{fig:J0} and \ref{fig:J025}, and test data for Fig.~\ref{fig:J0}.
It is symbolized in Fig.~\ref{fig:intro}(b) with the arrow (1), and the values of corresponding order parameter $O_{\text{CDW-I}}$ are plotted in Fig.~\ref{fig:OP}(a).
The transition, defined as above, occurs for $V_1/J=1$.
The second transition line is calculated for $V_2=0.25 \, V_1$ and $V_1/J = \la 0, 80 \ra$.
Indicated with the arrow (2), it is the source of test data for Fig.~3 of the main manuscript.
We plot the corresponding order parameter CDW-I in Fig.~\ref{fig:OP}(b), and the transition takes place for $V_1/J=1.85$.
The final transition line cuts three phases: LL, BO, and CDW-II.
It is marked with the arrow (3) and provides both training and test data for Fig.~4 of the main manuscript.
It is calculated for constant $V_1=1/J$ and $V_2 = \la 0, 8 \ra \, V_1$.
Transition between LL and BO occurs for $V_2 = 0.51 \, V_1$, and between BO and CDW-II for $V_2 = 1.7 \, V_1$.
It is important to notice that for the chosen range of parameters $V_2 = \la 1.7, 8 \ra \, V_1$, two phases co-exist what can be seen in Fig.~\ref{fig:OP}(c).

\section{Convolutional neural network}
\label{app:CNN}

\begin{figure}[h]
\begin{center}
\includegraphics[width=\columnwidth]{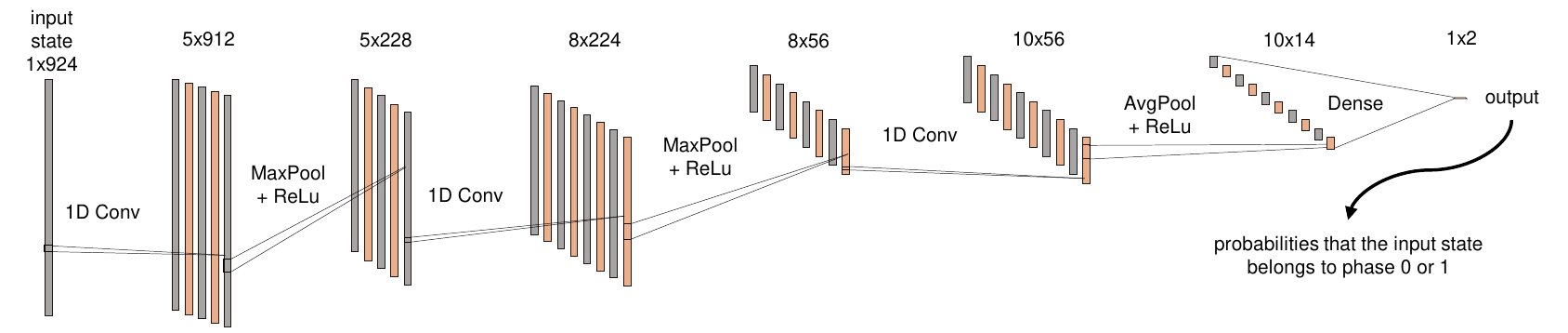}
\end{center}\vspace{-0.6cm}
\caption{Scheme of used architecture. Length scale does not apply.}
\label{fig:CNN}
\end{figure}

We use a neural network (NN) (see Fig. \ref{fig:CNN}) consisting of 3 one-dimensional convolutional layers with five filters on the input vector, eight filters on the first hidden layer, and ten filters for the last convolution layer.
After the first two convolutions, we apply a max-pooling layer to reduce the dimension, and the last convolutional layer is followed by an average pooling layer.
Finally, we have one fully connected layer with two output neurons that predict the labels.
When designing the architecture, we make sure that the convolutional part contains a large part of the NN's parameters.
For the training of the NN, we use state vectors from each phase as input and label them with 0 or 1 for each phase.
We obtain the state vectors via the exact diagonalization of the Hamiltonian \ref{eq:ham}.

We use $L_2$ regularization during the training to effectively decrease the certainty of the NN's predictions.
The undertrained NN with imperfect accuracy can provide better intuition behind the problem than overtrained one, whose predictions are impacted by overfitting.
Used CNNs had accuracy between 89 and 96\%.

\end{document}